\input harvmac 

\def\cA{ {\cal{A}} }  


\lref\connes{ 
A. Connes,  ``Non-commutative geometry, '' Academic Press 1994 }

\lref\maj{ S. Majid, ``Foundations of quantum group theory'' CUP 1995 }

\lref\malstrom{ 
J.~M.~Maldacena and A.~Strominger,
JHEP {\bf 9812}, 005 (1998)
[arXiv:hep-th/9804085].
}

\lref\mst{
J.~McGreevy, L.~Susskind and N.~Toumbas,
JHEP {\bf 0006}, 008 (2000)
[arXiv:hep-th/0003075].
}

\lref\GomisWI{
J.~Gomis, S.~Moriyama and J.~w.~Park,
arXiv:hep-th/0210153.
}

\lref\malda{
J.~M.~Maldacena,
Adv.\ Theor.\ Math.\ Phys.\  {\bf 2}, 231 (1998)
[Int.\ J.\ Theor.\ Phys.\  {\bf 38}, 1113 (1999)]
[arXiv:hep-th/9711200].
}

\lref\excor{
S.~Corley, A.~Jevicki and S.~Ramgoolam,
Adv.\ Theor.\ Math.\ Phys.\  {\bf 5}, 809 (2002)
[arXiv:hep-th/0111222].
}

\lref\finfac{
S.~Corley and S.~Ramgoolam,
Nucl.\ Phys.\ B {\bf 641}, 131 (2002)
[arXiv:hep-th/0205221].
}

\lref\mjr{
R.~de Mello Koch, A.~Jevicki and J.~P.~Rodrigues,
arXiv:hep-th/0209155.
}

\lref\bmn{
D.~Berenstein, J.~M.~Maldacena and H.~Nastase,
JHEP {\bf 0204}, 013 (2002)
[arXiv:hep-th/0202021].
}

\lref\constab{
N.~R.~Constable, D.~Z.~Freedman, M.~Headrick, S.~Minwalla, L.~Motl, 
A.~Postnikov and W.~Skiba,
JHEP {\bf 0207}, 017 (2002)
[arXiv:hep-th/0205089].
}

\lref\kpss{
C.~Kristjansen, J.~Plefka, G.~W.~Semenoff and M.~Staudacher,
strings,''
Nucl.\ Phys.\ B {\bf 643}, 3 (2002)
[arXiv:hep-th/0205033].
}

\lref\vaverl{
D.~Vaman and H.~Verlinde,
arXiv:hep-th/0209215.
}

\lref\verl{
H.~Verlinde,
arXiv:hep-th/0206059.
}

\lref\pasqsal{
V.~Pasquier and H.~Saleur,
Nucl.\ Phys.\ B {\bf 330}, 523 (1990).
}

\lref\mose{
G.~W.~Moore and N.~Seiberg,
Nucl.\ Phys.\ B {\bf 313}, 16 (1989).
}

\lref\alvgom{
L.~Alvarez-Gaume, C.~Gomez and G.~Sierra,
Nucl.\ Phys.\ B {\bf 319}, 155 (1989).
}

\lref\etvar{ Pavel Etingof, Alexander Varchenko, 
 math.QA/9907181  } 

\lref\witcs{
E.~Witten,
Commun.\ Math.\ Phys.\  {\bf 121}, 351 (1989).
}

\lref\reshtu{ N. Reshetikhin, VG Turaev, 
       Invent Math 103 (3): 547-597 1991  } 

\lref\mets{
R.~R.~Metsaev,
Nucl.\ Phys.\ B {\bf 625}, 70 (2002)
[arXiv:hep-th/0112044].
}

\lref\bn{
D.~Berenstein and H.~Nastase,
arXiv:hep-th/0205048.
}

\lref\metsey{
R.~R.~Metsaev and A.~A.~Tseytlin,
Phys.\ Rev.\ D {\bf 65}, 126004 (2002)
[arXiv:hep-th/0202109].
} 

\lref\charpress{ V. Chari and A. Pressley,
 ``A guide to quantum groups, '' CUP 1994  } 
\lref\jr{
A.~Jevicki and S.~Ramgoolam,
JHEP {\bf 9904}, 032 (1999)
[arXiv:hep-th/9902059].
}
\lref\hrt{
P.~M.~Ho, S.~Ramgoolam and R.~Tatar,
Nucl.\ Phys.\ B {\bf 573}, 364 (2000)
[arXiv:hep-th/9907145].
}

\lref\conmos{ A. Connes and H. Moscovici, ``Cyclic cohomology, Hopf
Algebras and the Modular Theory, '' Math.QA/9905013 } 

\lref\quesne{ C. Quesne, 
J. Phys. A: Math. Gen. 25 ( 1992 ) 5977-5998 } 

\lref\gmr{
D.~J.~Gross, A.~Mikhailov and R.~Roiban,
arXiv:hep-th/0208231.
}

\lref\hssv{
Y.~H.~He, J.~H.~Schwarz, M.~Spradlin and A.~Volovich,
arXiv:hep-th/0211198.
}

\lref\ckt{
C.~S.~Chu, V.~V.~Khoze and G.~Travaglini,
JHEP {\bf 0206}, 011 (2002)
[arXiv:hep-th/0206005].
}

\lref\schnizer{ W.A. Schnizer, J. Math. Phys. { \bf 34 }, 1993 }

\lref\cuza{
T.~L.~Curtright, G.~I.~Ghandour and C.~K.~Zachos,
J.\ Math.\ Phys.\  {\bf 32}, 676 (1991).
}

\lref\gurone{
U.~Gursoy,
arXiv:hep-th/0208041.
}

\lref\ckprt{
C.~S.~Chu, V.~V.~Khoze, M.~Petrini, R.~Russo and A.~Tanzini,
arXiv:hep-th/0208148.
}

\lref\sv{
M.~Spradlin and A.~Volovich,
Phys.\ Rev.\ D {\bf 66}, 086004 (2002)
[arXiv:hep-th/0204146].
}

\lref\Gopakumar{
R.~Gopakumar,
arXiv:hep-th/0205174.
}

\lref\kklp{
Y.~j.~Kiem, Y.~b.~Kim, S.~m.~Lee and J.~m.~Park,
arXiv:hep-th/0205279.
}

\lref\huang{
M.~x.~Huang,
Phys.\ Lett.\ B {\bf 542}, 255 (2002)
[arXiv:hep-th/0205311].
}

\lref\lmp{
P.~Lee, S.~Moriyama and J.~w.~Park,
arXiv:hep-th/0206065.
}

\lref\svtwo{ 
M.~Spradlin and A.~Volovich,
arXiv:hep-th/0206073.
}

\lref\ksv{
I.~R.~Klebanov, M.~Spradlin and A.~Volovich,
Phys.\ Lett.\ B {\bf 548}, 111 (2002)
[arXiv:hep-th/0206221].
}

\lref\bers{
M.~Bianchi, B.~Eden, G.~Rossi and Y.~S.~Stanev,
arXiv:hep-th/0205321.
}

\lref\bkpss{
N.~Beisert, C.~Kristjansen, J.~Plefka, G.~W.~Semenoff and M.~Staudacher,
arXiv:hep-th/0208178.
}

\lref\cfhm{
N.~R.~Constable, D.~Z.~Freedman, M.~Headrick and S.~Minwalla,
JHEP {\bf 0210}, 068 (2002)
[arXiv:hep-th/0209002].
}

\lref\psvvv{
J.~Pearson, M.~Spradlin, D.~Vaman, H.~Verlinde and A.~Volovich,
arXiv:hep-th/0210102.
}

\lref\rsv{
R.~Roiban, M.~Spradlin and A.~Volovich,
arXiv:hep-th/0211220.
}

\lref\gursoytwo{
U.~Gursoy,
arXiv:hep-th/0212118.
}

\lref\dg{
S.~R.~Das and C.~Gomez,
JHEP {\bf 0207}, 016 (2002)
[arXiv:hep-th/0206062].
}
\lref\dgr{
S.~R.~Das, C.~Gomez and S.~J.~Rey,
Phys.\ Rev.\ D {\bf 66}, 046002 (2002)
[arXiv:hep-th/0203164].
}

\lref\oku{
K.~Okuyama,
JHEP {\bf 0211}, 043 (2002)
[arXiv:hep-th/0207067].
}

\lref\BlauMW{
M.~Blau, J.~Figueroa-O'Farrill and G.~Papadopoulos,
Class.\ Quant.\ Grav.\  {\bf 19}, 4753 (2002)
[arXiv:hep-th/0202111].
}

\lref\BlauDY{
M.~Blau, J.~Figueroa-O'Farrill, C.~Hull and G.~Papadopoulos,
Class.\ Quant.\ Grav.\  {\bf 19}, L87 (2002)
[arXiv:hep-th/0201081].
}

\lref\bgm{
P.~Berglund, E.~G.~Gimon and D.~Minic,
JHEP {\bf 9907}, 025 (1999)
[arXiv:hep-th/9905097].
}

\Title{ \vbox{\baselineskip12pt\hbox{  Brown HET-1336 }}}
 {\vbox{
\centerline{     Strings on  Plane Waves, }
\vskip.08in
\centerline{   Super-Yang Mills in Four Dimensions, }
\vskip.08in
\centerline{    Quantum Groups  at Roots of One } }}

\vskip.08in

\centerline{$\quad$ { Steve Corley, Sanjaye Ramgoolam   } }
\smallskip
\centerline{{\sl Department of Physics}}
\centerline{{\sl Brown  University}}
\centerline{{\sl Providence, RI 02912 }}
\smallskip
\centerline{{\tt scorley@het.brown.edu }}
\centerline{{ \tt ramgosk@het.brown.edu }}

\vskip .3in 

We show that the BMN operators in $D=4$ $N=4$ super Yang Mills theory 
proposed as duals of stringy oscillators in a 
plane wave background have a natural 
quantum group construction in terms of the 
quantum deformation of the  $SO(6)$  $R$ symmetry. 
We describe in detail how a $q$-deformed $U(2)$  
subalgebra generates 
BMN operators, with $ q \sim e^{ 2 i \pi \over J }$.
The standard quantum co-product 
as well as generalized traces which 
use  $q$-cyclic operators acting on tensor products 
of Higgs fields are the ingredients in this construction.  
They  generate the oscillators with the correct 
( undeformed )  permutation symmetries of Fock space oscillators.  
The quantum group can be viewed as a spectrum generating algebra, 
and suggests that correlators of BMN operators should have 
a geometrical meaning in terms  of spaces with quantum group symmetry.

\Date{ December  2002 } 

\def\hovt{ { H \over 2 } }
\def\ovt{  \over 2  }

\def\di{ \partial } 
\def\diz#1{ \partial z_{#1 }} 
\def\dizb#1{ \partial \bar z_{#1} } 
\def\zb#1{ \bar z_{#1} } 
\def\cS{ { \cal { S }} }  
\def\cF{ { \cal { F }} }  
\def\cT{ { \cal { T }} }  
\def\kt{ {\tilde k} }
\def\Pd{ \Phi^{\dagger}}

\newsec{ Introduction } 

 Type IIB String theory on  a plane wave background with RR flux 
 has recently been discovered to be solvable 
 \refs{\mets, \metsey }. This background is a 
 limit of the $ ADS_5 \times S^5 $ spacetime, and  
 a gauge theory dual has been proposed
 by Berenstein, Maldacena, Nastase (BMN) \bmn. 
 The proposal builds on the ADS/CFT duality \malda\ 
 between type IIB string theory on $ADS_5 \times S^5 $
 and $N=4$ super-Yang Mills theory in four dimensions. 
 BMN identified the operator $ tr ( \Phi_1^L  ) $ 
 on the gauge theory side as corresponding to the 
 vacuum of the string theory on the 
  plane wave background. Here  $ \Phi_1$ is 
 a complex Higgs field obtained from combining two 
 hermitian Higgs fields chosen from the six appearing in 
 the super-Yang Mills. 
 Modifications  of this operator are obtained by inserting, 
 with some phase factors,   other fields in 
 the $SU(4 | 2,2)$ multiplet of the  gauge theory. 
 The $ SU(4) \sim SO(6)$ subalgebra allows the insertion 
 of other Higgs fields. The phase factors are of the 
 form $ e^{ 2 \pi i p   \over J } $ where $p$ is a momentum carried 
 by the corresponding stringy oscillator and $ J = L-n$, 
$n$ being the number of impurities. Correlation functions of 
these operators in the gauge theory and the comparison 
with string theory have been discussed in many papers, 
see for example 
\refs{\bmn\sv\kpss\bn\constab
\Gopakumar\kklp\huang\bers\ckt\lmp\svtwo\ksv\gurone
\ckprt\bkpss\gmr\cfhm\mjr\psvvv\GomisWI\hssv\gursoytwo-\rsv }. 
Discussions of the symmetries of this ppwave background 
have appeared in \refs{\BlauDY ,\BlauMW \dgr, \dg, \oku }

 In this paper we start by observing that 
 the construction of the BMN operator 
 involving a set of impurities all of  the 
 same  momentum $p=-1$  can be viewed as the result 
 of the standard action of a generator of $SO_q(6)$  
 on the $J +n $ fold tensor product of
 Higgs fields, followed by a trace. 
 The $q$-deformed action on tensor products 
 follows from the quantum co-product which is 
 necessary if we want the quantum group relations 
 to be preserved on the tensor product. In other words, 
 choosing the  linear combinations of  
 impurity insertions weighted by $q$ factors is 
 equivalent to choosing a set of states which, 
 along with the vacuum, form a representation of the 
 quantum group. 
 The $q$-deformation 
 parameter is $ q = e^{ 2 \pi i    \over J } $. 
 For concreteness we describe this in detail for  the case 
 where the impurity is another complex Higgs 
 say $ \Phi_2$. For these insertions, we only need 
 a $q$-deformed $U(2)$ or   $U_q ( U(2) ) $. Relevant facts about $U_q
 ( U(2) )  $  
 are recalled in section 3 and this simplest quantum group 
 construction of a BMN operator is described 
 in section 5.1. This suggests that we should 
 view the quantum group  $U_q (U(2) ) $, and more generally 
 $ SU_q ( 4|2,2 )$ ), for $ q = e^{ 2 \pi i    \over J } $ 
 as a spectrum generating  algebra for BMN operators. 

 A superficial look at phases $ e^{ 2 \pi i p \over J } $ 
 which enter  the construction of BMN operators corresponding to 
 stringy oscillators with generic momenta   
 would suggest that a quantum group $ SU_q ( 4|2,2 )$ depending 
 on a single parameter $q= e^{ 2 \pi i    \over J } $  would not have enough 
 structure to give the general BMN operator.  One of main
points of this paper is to show that generic momenta
are nevertheless obtained for a single $q$ deformation parameter 
 in $ SU_q ( 4|2,2 )$.

 Physical applications of quantum groups in two-dimensional 
 CFT and three dimensional Chern Simons theory 
 \refs{ \pasqsal, \mose , \witcs, \alvgom, \reshtu }  show that, 
 in addition to the quantum co-product an interesting 
 role is played by quantum traces. 
 In the Mathematics literature quantum traces 
 with different choices of Cartan elements 
 have been shown to have interesting properties \etvar. 
 Further, non-commutative geometry of spaces with 
 quantum group symmetry, requires the use generalized traces \conmos.    
 These observations suggest that we should look for a construction 
 of BMN operators involving the quantum group 
 $ SU_q ( 4|2,2 )$ with fixed $q$, and hence fixed 
 quantum co-product, but using constructions 
 which can be viewed as generalized traces. 

 Our generalized traces are constructed by 
 composing the action of the co-product, 
 with some $q$-cyclic operators, denoted by 
 $ \tau $  and then taking 
 a trace. The $q$-cyclic operators act on a tensor 
 product of Higgs fields and produce a sum 
 of tensor products of Higgs fields, where each successive 
 term in the sum involves a cycling of the Higgs 
 fields accompanied by an additional phase factor  
 which depends on the weight of the Higgs being cycled 
 under a choice of element in the Cartan of $ SU_q ( 4 | 2,2 )$.   
 These $\tau $ operators are defined carefully in section 4. 
 Our concrete calculations are done for  
 $ U_q( U(2) ) $ but the  main ideas generalize to the 
 full superalgebra. 

 An important property of the quantum group construction is  
 that it automatically produces BMN operators with the 
 correct permutation symmetries. Since they are 
 dual to string theory oscillators which commute with each other, 
 they should have the corresponding permutation symmetries. 
 These symmetries have been discussed with some care 
 in \refs { \constab ,  \mjr }. Note that we are here 
 focusing on a $U(2)$ subalgebra of $ SU(4|2,2) $ which produces 
 bosonic oscillators. More generally there will be fermionic
 oscillators which will involve anti-commutation properties.    
 We review these symmetric BMN operators in section 2, 
 and some other details about their properties are 
 described in Appendix A. Our main technical result 
 is that the 
 $U_q ( U(2) ) $ quantum group construction, with 
 a single value of $q$, using the standard co-product, 
 along with a sequence of $q$-cyclic ( $\tau $ ) operators 
 followed by a trace, automatically  reproduces all stringy 
 states with a fixed number of string oscillators, obeying 
 the correct permutation symmetry.  
 The different $q$ we need for different numbers 
 of impurities differ by factors of $ 1/J $, where 
 $J$ is large in the BMN limit, so in effect 
 all symmetric BMN operators involving a single 
 impurity type are produced by the $U_q ( U(2) ) $ 
 quantum group construction, with a fixed $q$. 
 The construction of states 
 involving general momenta using the co-product and 
 generalized traces is described in section 5. 
 
 In section 6, 
 we  outline how our construction of BMN operators 
 can lead to formulae for correlators as traces 
 of quantum group operators in tensor spaces, generalizing 
 the work of  \refs{ \excor , \finfac  } where traces of 
 projectors of classical groups in tensor spaces were 
 related to correlators of SYM. We outline how the 
 spectrum-generating quantum algebra acts on the 
 super-Yang Mills action, showing that  its action can be given a
 well-defined meaning but that, as expected, the SYM action is not
 invariant.  We discuss the 
 geometrical meaning of our algebraic quantum group 
 construction of BMN operators in terms of quantum spaces, 
 by using similarities  between 
 the $\tau$ we have used and some analogous operators 
 that appear in the cyclic cohomology of quantum groups.

\newsec{ Review of BMN operators } 

We begin by reviewing the BMN correspondence between large R-charge
operators of the ${\cal N}=4$ super-Yang-Mills (SYM) theory
and string states in a pp-wave background.  The SYM theory
contains six real scalar fields $X^i$ where $i=1,...,6$.  To
express the theory in ${\cal N}=1$ notation, these are combined
into three complex combinations $\Phi_j = X^j + i X^{j+3}$
where $j=1,2,3$.  The theory contains an $SU(4)$ R-symmetry
( subgroup of the $SU(4|2,2)$ superalgebra symmetry )
under which the scalars transform in the six-dimensional
representation.  To construct the BMN operators, one selects
a $U(1)$ subgroup of the R-symmetry group, or equivalently
chooses one of the complex scalars as the ``background''
scalar.  For example, selecting the $\Phi_1$ scalar, then the
ground state of the string theory in a pp-wave background
$| 0 \rangle$ corresponds to the operator
\eqn\ground{ | 0 \rangle \leftrightarrow {\cal N}_J ~
TR ( \Phi_{1}^J) }
where the R-charge $J$ of the operator corresponds to the light
cone momentum $p^+$ on the string theory side and the
factor ${\cal N}_J$ is a normalization factor which will
not be important for our purposes here.

Exctations above  the ground state arise on the SYM side
by inserting the other scalars $\Phi_2$ and $\Phi_3$ and
their complex conjugates (there are other possibilities
as well, generated by the superalgebra, 
 that are in fact necessary to match with the 
string theory states, but as we shall not be discussing
these other cases we refer the reader to the literature
for more details, see eg. \bmn).   Moreover these ``impurities'' are 
accompanied by phase factors whose powers correspond to
oscillator number on the string theory side.  Appendix A
is devoted to a careful exposition
of these operators.  For those not interested in the details
however we shall simply record the final result here.  The BMN operator
with $\Phi_{\beta_l}$ insertions, with momenta $p_l$, 
 for $1 \leq l \leq n$ is given
by
\eqn\BMNfinal{\eqalign{
& {\cal O}_{  \beta_n,p_n ;   \beta_1,p_1,
\beta_2,p_2, \cdots \beta_{n-1}, p_{n-1}} \cr 
& \qquad =  {\cal N}^{\prime}_n
\sum_{0 \leq k_1 \leq k_2 \leq \cdots \leq k_{n-1} \leq J}
~ \sum_{\tau \in S_{n-1}} q^{\sum_{l=1}^{n-1} k_l p_{\tau (l)}} \cr 
& \qquad \qquad \qquad
 TR  [ {\beta_n \atop ,} k_1 {\beta_{\tau (1)} \atop ,} k_2 - k_1
{\beta_{\tau (2)} \atop ,} \cdots 
  {\beta_{\tau (n-1)} \atop ,}
J - k_{n-1} ] \cr }}
where we have introduced the convenient notation
\eqn\opnotation{ TR  \left(\Phi_{1}^{k_1} \Phi_{\beta_1}
\Phi_{1}^{k_2} \Phi_{\beta_2} \cdots \Phi_{1}^{k_n} \Phi_{\beta_n}
\Phi_{1}^{k_{n+1}} \right) \equiv TR  [ k_1 {\beta_1 \atop ,}
k_2 {\beta_2 \atop ,} \cdots k_n {\beta_n \atop ,} k_{n+1} ] }
where $\beta_i = 2,3$.
In words, traces of products of operators are denoted
as above with commas corresponding to impurities and
the number above the comma indicating the type of impurity.
The integers between the commas indicate the power
of the background field.  We have given the BMN operator
for $\Phi_2$ and $\Phi_3$ impurities, but the extension
to the complex conjugate fields is trivial.  The sum on $\tau$ is over
the permutation group $S_{n-1}$.
We will often be interested in the case where 
all the impurities are a complex $ \Phi_2$. 
In this case, we can drop the $\beta $ labels
 and just write 
\eqn\opnoti{ TR  \left(\Phi_{1}^{k_1} \Phi_2
\Phi_{1}^{k_2} \Phi_2 \cdots \Phi_{1}^{k_n} \Phi_2
\Phi_{1}^{k_{n+1}} \right) \equiv TR  [ k_1 ,
k_2 , \cdots k_n , k_{n+1} ]}
where now the commas are assumed to always correspond
to $\Phi_2$ impurities.

Via the BMN correspondence this operator corresponds on the
string side to the state
\eqn\BMNdual{
{\cal O}_{  \beta_n,p_n ;   \beta_1,p_1,
\beta_2,p_2, \cdots \beta_{n-1}, p_{n-1}}
\leftrightarrow \prod_{l=1}^{n} \alpha^{\dagger ~\beta_l }_{p_l} 
| 0 \rangle .}
The momenta $p_l$, for $l=1 \cdots n-1$ appear explicitly in 
\BMNfinal\ and $p_n$ is fixed by the constraint
$p_1 + p_2 + \cdots p_n = 0$ as follows from reparametrization
invariance of the string worldsheet. The treatment of  
  $p_n$ in \BMNfinal\ appears to break the $S_n$ symmetry 
 of the state in \BMNdual\ but cyclicity together with the 
 condition $q^J=1$ implies that it does not. 
  We would like to
point out that while $p_1 + p_2 + \cdots p_n = 0$ is
necessary for the correspondence \BMNdual\ to make
sense, it is possible to generalize the
operator 
${\cal O}_{\beta_n,p_n;\beta_1, p_1, ...;\beta_{n-1}, p_{n-1} }$ to
the case where $p_1 + p_2 + \cdots p_n \neq 0$.
The correspondence \BMNdual\ is then modified
so that ${\cal O}_{\beta_n,p_n;\beta_1, p_1, ...;\beta_{n-1}, p_{n-1}
}$ corresponds
to a linear superposition of single string states.
This is discussed in detail in Appendix A.

In the case of just $\Phi_2$ insertions, 
\BMNfinal\ simplifies to
\eqn\BMNtwo{\eqalign{
& {\cal O}_{p_n; p_1... p_{n-1} } = \cr 
&  {\cal N}^{\prime}_n
\sum_{0 \leq k_1 \leq k_2 \leq \cdots \leq k_{n-1} \leq J}
~ \left( \sum_{\tau \in S_{n-1}} q^{\sum_{l=1}^{n-1} k_l p_{\tau (l)}}
\right)  TR  [, k_1 , k_2 - k_1  
, \cdots , J - k_{n-1} ] \cr }}
In particular in the sum over permutations,  $\tau$
only  enters the $q$-factor and not the operator. 
This is the form of the operator that we will be comparing
to later.  To get some feel for these operators we
give some examples.  If all $p_l$'s for $1 \leq l \leq n-1$
are equal, then the sum over permutations reduces to
just one term, i.e., the $q$-factor becomes
$q^{p(k_1 + \cdots k_{n-1})}$ with $p$ denoting
the common values of the $p_l$'s.  The correspondence
\BMNdual\ then becomes
\eqn\BMNdualone{{\cal O}_{-(n-1)p; p, \dots p}
\leftrightarrow  \alpha^{\dagger}_{-(n-1) p} (\alpha^{\dagger}_p)^{n-1} 
| 0 \rangle.}
A slightly more non-trivial case is to let
$p_1 = \cdots = p_{n-2} = p \neq p_{n-1}$.
The sum over permutations of the phase factor in \BMNtwo\ then
reduces to 
\eqn\phaseeg{q^{p (k_1 + \cdots + k_{n-1})} \sum_{l=1}^{n-1}
q^{(p_{n-1} - p) k_l}.}
Moreover the correspondence \BMNdual\ reduces to
\eqn\BMNdualtwo{ {\cal O}_{-(n-2)p - p_{n-1};  p, \dots ,p,  p_{n-1}}
\leftrightarrow  \alpha^{\dagger}_{-(n-2) p - p_{n-1}} 
(\alpha^{\dagger}_p)^{n-2} \alpha^{\dagger}_{p_{n-1}} 
| 0 \rangle.}
A more detailed discussion of the BMN
operators appears in Appendix A.

\newsec{ Review of quantum group facts } 

We begin by reviewing some facts about quantum  algebras, 
focusing on the quantum deformation of $SU(2)$, which is 
the non-trivially deformed part of $U_q (U(2) ) $. 
For more details see for example \charpress\maj.   
The quantum algebra $U_q( SU(2) )  $ is generated by $ H, X_+, X_- $
with relations : 
\eqn\relqsut{\eqalign{ 
&  [ H, X_{\pm } ] = \pm  2 X_{\pm } \cr 
&   [ X_{+} , X_{-} ] = { q^H - q^{-H} \over q - q^{-1} } \cr 
}}
In the limit $ q \rightarrow 1 $, this approaches 
the classical algebra : 
\eqn\relcsut{\eqalign{ 
&  [ H, X_{\pm } ] = \pm 2 X_{\pm } \cr 
&   [ X_{+} , X_{-} ] =  H \cr 
}}

An important property which the quantum algebra 
shares with the classical algebra, is that if 
$ X_{\pm} , H $ are represented as operators 
acting on $V_1$  and $V_2$ obeying the quantum relations, then 
$ V_1 \otimes V_2$ is also a representation. This is 
only true, however if the quantum group generators
are taken to act on the tensor product using the 
quantum co-product. The quantum co-product 
can be viewed as a map $ \Delta $  from the algebra $U_q$
to $ U_q \otimes U_q$ 
\eqn\copform{\eqalign{  
& \Delta ( H ) = H \otimes  1 + 1 \otimes H  \cr 
&  \Delta ( q^H ) = q^H \otimes q^H \cr  
& \Delta ( X_{\pm} ) = X_{\pm} \otimes q^{H\ovt } + 
q^{-H \ovt } \otimes X_{\pm }
\cr 
}}
One can check for example that 
\eqn\arel{ 
[ \Delta ( H ) ,\Delta (  X_{\pm }) ] =  \pm 2 \Delta ( X_{\pm } ) 
}
Equivalently we may write 
\eqn\equvw{\eqalign{  
&  q^H X_{\pm} q^{-H} = q^{ \pm 2}  X_{\pm} \cr 
&  \Delta ( q^H )  X_{\pm } \Delta ( q^{-H} )  =  q^{\pm 2}  
\Delta ( X_{\pm } ) \cr }}
 In the limit $ q \rightarrow 1$ the quantum co-product 
leads to the ordinary action of the algebra on the 
tensor products. 

An important point worth  noting is that, given the normalizations 
used in \relcsut\ the eigenvalues of $H$ in the fundamental 
representation are $ 1$ and $-1$. On the state with 
$ H=1 $, we have $ { q^{H} - q^{-H} \over q - q^{-1} } = 1 = H  $. 
On the state with $H=-1$, we have  
$ { q^{H} - q^{-H} \over q - q^{-1} } = -1 = H  $. 
This means that the matrices representing $U_q SU(2) $ 
in the fundamental representation are the same as the 
ones representing the classical $SU(2)$. 
This fact is quite general, see for example the case 
of $SO_q(2n)$,  which includes the $SO_q(6)$ of interest here, 
in \schnizer. 
Finite dimensional 
representations can be constructed from the tensor products
of the fundamental one. The matrices in these tensor 
products differ because of the different co-products. 
In a sense, as far as finite dimensional 
representations are concerned,  the essence of the quantum deformation 
is in the quantum co-product. The close relation between 
finite dimensional representations of the quantum group 
and the classical group is discussed in the physics 
literature in  \cuza.


\subsec{  Quantum co-product }

Using the co-product \copform\  we can consider the action 
on $ V \otimes V$ which we denote $ \Delta_2$.   
\eqn\copslt{ \Delta_2 ( X_+ ) = X_+ \otimes q^\hovt + q^{-\hovt} \otimes X_+ } 
Now consider the action of $X_+ $ 
on a tensor product of three vector spaces.
We can think of $ ( V \otimes V \otimes V ) $ 
as $ ( V \otimes V ) \otimes V $.     
\eqn\copthr{\eqalign{ 
&  \Delta_3 ( X_+ ) = \Delta_2 ( X_+ ) \otimes q^\hovt 
                                  + \Delta_2 (q^{-\hovt} ) \otimes    X_+
 \cr 
& = X_+ \otimes  q^\hovt \otimes q^\hovt  + q^{-\hovt} \otimes X_+ \otimes q^\hovt 
    + q^{-\hovt} \otimes q^{-\hovt} \otimes  X_+ \cr }}
By an induction argument, we can show that 
$ \Delta_n ( X_{+} ) $ is a sum where 
$X_+$ acts successively on each of the $n$ factors, 
while $q^{ - H \ovt } $ acts on  the factors to the left  
and $ q^{ H \ovt } $ acts on the factors to the right. 
For the future use, we will express this by denoting the 
action of any generator $ X$  on the $ k$'th factor 
as $ \rho_k (X )$. We have the formula 
\eqn\formula{ 
\Delta_n ( X_{\pm}  ) = \sum_{ k=1}^n  \rho_k ( X_{\pm} ) 
  \bigl ( \sum_{l=1}^{ k-1 } \rho_l ( q^{ - H \ovt } )     + \sum_{l=
k+1}^{n} \rho_l ( q^{ H \ovt } )    \bigr )  }

If we are considering the action on 
states where $H=1$ this leads to the weighting 
of the action of $X_+$ by a power of $q$ which depends 
on where the $X_+$ is acting.  Acting on $V^{\otimes 3}$
for example, one gets the sequence of $q$-factors
$q, 1, q^{-1} $.  Acting on 
$V^{\otimes 4 } $, we get weights $ (q^{3 \over 2} , q^{1 \over 2} , 
q^{-1 \over 2 } , q^{-3\over 2} ) $. 
More generally, when $ \Delta_n$  is acting on a product of states where 
$H=1$ we get 
\eqn\resi{
\Delta_n ( X_{\pm}  ) =  \sum_{k=1}^{n} 
q^{n + 1 \over 2 } q^{-k}    \rho_k ( X_\pm  )   }

\newsec{Embedding of $U(2)$ in $SO(6) $ and the $q$-cyclic operations }

We  describe with  the 
$SO(6) $ algebra and 
and the relevant $U(2)$ subgroup 
which will be deformed according to the 
formulae in section 3.. 

Take the standard action of $SO(6)$ on 
$ x_1, \cdots x_6 $. Let us form 
combinations
\eqn\combs{\eqalign{ 
&  z_1 = x_1 + i x_4   \cr 
& z_2 = x_2 + i x_5 \cr 
& z_3 = x_3 + i x_6 \cr }}

The Cartan subalgebra  is spanned by 
\eqn\cart{\eqalign{ 
&  H_1  = z_1 {\di \over \diz{1} }   - \zb{1}{ \di \over \dizb{1}}  \cr  
&   H_2  = z_2 {\di \over \diz{2} }   - \zb{2}{ \di \over \dizb{2 } }  \cr 
&  H_3  = z_3 {\di \over \diz{3} }   - \zb{3}{   \di \over \dizb{3} } \cr }} 

Additional  generators of the $SO(6)$ Lie algebra 
are, for $i \ne j $ running from $1$ to $3$ :  
\eqn\lieal{
E_{ij} = z_i { \di \over \diz{j}   } -
 \zb{j} { \di \over \dizb{i} } } 
We also take, for $ i < j $, 
\eqn\ilej{\eqalign{  
&  P_{ij} =   z_i { \di \over \dizb{j}   } - z_j { \di \over \dizb{i}
 } \cr 
& Q_{ij} = 
 - \zb{i} { \di \over \diz{j}   } + \zb{j} { \di \over \diz{i} }
\cr 
}}
It is easy to check that the above
operators preserve $ z_1 \zb{1} +  z_2 \zb{2} +  z_3 \zb{3} $. 
and that 
\eqn\getzth{\eqalign{ 
& E_{21} ( z_1 ) = z_{ 2 } \cr 
& E_{21} ( z_{2} ) = 0 \cr 
}}

In the context of maximally supersymmetric  SYM with 
$U(N)$ gauge group,  these generators 
act on complex fields  $\Phi_1, \Phi_2, \Phi_3$, which 
are matrices of size $N$,  instead of
complex variables $ z_1, z_2, z_3$. 
For the dual  of string theory on 
 plane waves \bmn\ we are interested in 
operators which are close to $ tr ( \Phi_1^ J ) $ for some large $J
\sim { \sqrt N }$. The action of  $SO(6) \sim SU(4)  $ 
can be used to generate insertions of other  scalar 
operators. Note that if we use the full 
$SO(6)$ algebra we would generate insertions of 
$ \Phi_2, \Phi_3, \Phi_2^{ \dagger} ,  \Phi_3^{ \dagger}  $ 
as well as $ \Phi_1^{ \dagger} $. Operators which include insertions
of $ \Phi_1^{\dagger} $ are actually not of interest 
in the BMN limit, because strong coupling effects 
give them infinite dimensions. 
If we work with $ U(3)$ generated by $ H_1, H_2, H_3, E_{21} , E_{12}
, E_{32} , E_{23} $ we can get insertions of the holomorphic 
$ \Phi_2, \Phi_3$ but not the $ \Phi_2^{\dagger}, \Phi_3^{\dagger}$
nor the $ \Phi_1^{\dagger}$. The supersymmetric version of this 
will be a superalgebra $ SU(3| 2,1)$. 
In this paper we will 
focus on a $U(2) $ subgroup of this $SU(3)$ and describe in detail  
the connection between the $q$-deformed $U(2)$ and the 
BMN operators involving insertions of $\Phi_2$.

The $ SU(2) $ subgroup of interest is generated by 
$ E_{12} = X_+ , X_- = E_{21}$ and $H =  H_1 - H_2 $
which obey the relations \relcsut. The extra $U(1)$ which gives 
$U(2)$ is generated by $H_1 + H_2$.
The quantum group relations are 
\relqsut. The co-products of the diagonal generators $H_1, H_2 $ 
are unchanged.

\subsec{$q$-cyclic operations } 

To construct BMN operators in the next section we
will require the use of a generalized trace which
we define in this section.  To construct this operator
consider the algebra $\cA$ of Higgs fields (for simplicity
just $\Phi_1$ and $\Phi_2$ for the purposes of this paper)
and its tensor products $\cA^{\otimes L}$ acted on
by the quantum group described in the previous section.
We define $ \tau_{( a, b ) } $ 
as a map from $ \cA^{\otimes L } $ to 
$ \cA^{\otimes L } $ by
\eqn\defntau{\tau_{(a,b)} (\Phi_{\beta_1} \otimes
\Phi_{\beta_2} \otimes \cdots \otimes \Phi_{\beta_L})
= \sum_{i=1}^{L} \Phi_{\beta_{i+1}} \otimes \cdots
\Phi_{\beta_L} \otimes \Delta_{i} (q^{a H_1 + b H_2}) 
(\Phi_{\beta_{1}} \otimes \cdots \otimes \Phi_{\beta_{i}} ),}
i.e., it is a sum of cyclic permutations of
$\Phi_{\beta_1} \otimes \cdots \otimes \Phi_{\beta_L}$
weighted by $q$-dependent factors.  The factor is easy
to determine because the $\Phi_k$'s are eigenstates of
the $H_i$'s with eigenvalues $\delta_{k,i}$.  Therefore
the weighting factor is simply $q^{a n_1 + b n_2}$
where $n_1$ and $n_2$ are the number of $\Phi_1$
and $\Phi_2$ fields cycled respectively.  As an example
consider the operator $ \Phi_1 \otimes \Phi_2 \otimes \Phi_2 $.
Applying $\tau_{(a,b)}$ we 
find
$$ \tau_{ ( a,b )}  ( \Phi_1 \otimes \Phi_2 \otimes \Phi_2 ) 
 =q^{a}  \Phi_2 \otimes \Phi_2 \otimes \Phi_1 
+ q^{a + b } \Phi_2 \otimes \Phi_1\otimes \Phi_2  
+q^{a + 2b }  \Phi_1 \otimes \Phi_2 \otimes \Phi_2  $$
Note that we have the relations
\eqn\qcyc{\eqalign{  
& \tau_{ ( a,b )} ( \Phi_1 \otimes \Phi_2 \otimes \Phi_2 ) \cr 
&  = q^{a} \tau_{ ( a,b )} ( \Phi_2 \otimes \Phi_2 \otimes \Phi_1 ) \cr 
& = q^{a+b} \tau_{ ( a,b )} ( \Phi_2 \otimes \Phi_1 \otimes \Phi_2  )
\cr
& = q^{a + 2b} \tau_{ ( a,b )} ( \Phi_1 \otimes \Phi_2 \otimes \Phi_2
) }}
as follows from the definition of $\tau_{(a,b)}$.
In other words, each time we cycle a Higgs field, 
we pick up a phase which is determined by the 
charge of the Higgs field under $ q^{a H_1 + bH_2 } $. 
For the first and last lines of \qcyc\ to be consistent,
we require that $q^{a + 2b} = 1$.  More generally,
given an element of the tensor product algebra with
$J$ $\Phi_1$'s and $n$ $\Phi_2$'s, we must demand
that $q^{Ja + nb}=1$.  Moreover if we demand that
$q^J = 1$, as will be done in this paper, then we 
must further require that
$b=0 (mod J)$.
If $a=b=0$ we have the standard 
cyclicity of traces, except that it is here expressed as a
 property of a map from $\cA^{\otimes L } $ to  $\cA^{\otimes L } $. 

In the next section it will be more convenient to rewrite
the q-cyclic operator defined above \defntau\ as
\eqn\formqcyc{ 
\tau_{ a, b } = \sum_{k=1}^{L}  c^k   q^{  \sum_{l=1}^{k} \rho_{l} ( aH_1 +
bH_2 )}  } 
 The operator $c$ cycles one Higgs field through the left. 
 The operator  $c^k$ has the effect of performing  a $k$-step 
 cycling operation. The sum over $l$ at fixed $k$ 
 is an instruction to pick up a factor of $q^{aH_1 +bH_2}  $ for each 
 Higgs field cycled.

\newsec{ The construction of the BMN operators } 

\subsec{ With coproduct and trace }  

   We will be interested in the action of 
  $ \Delta_q ( E_{21}^n   ) $ on  $ \Phi_1^{\otimes L  } $ 
  which will lead to BMN operators with $ J $ 
  $ \Phi_1$ operators ( where $ J = L-n $ )
   and $ n$ copies of the $ \Phi_2$ operator. 
  The simplest way to get a class of BMN operators from this 
  action of the quantum group is to multiply the $\Phi$'s in the 
  tensor product and then take a trace of the resulting matrix.  
 We will denote the result of this 
  combined multiplication and tracing operation
  applied to $ \Delta_q ( E_{21}^n   )  \Phi_1^{\otimes L  }$  as 
  $ TR ( \Delta_q ( E_{21}^n ) \Phi_1^{\otimes L  }  )$.  
 More general operators  can be obtained by 
 considering generalized traces such as the ones
defined by combining the ordinary trace
with the $q$-cyclic operators described in section 4.
Before analysing these more general cases, we shall
first consider just the ordinary trace.

To begin we derive a convenient expression for the
operator $ \Delta_q ( E_{21}^n )$ using its definition
given in \formula.  Substituting \formula\ into $ \Delta_q ( E_{21}^n )$
we obtain
\eqn\deltaq{\eqalign{  
& \Delta_q ( E_{21}^n ) = 
\sum_{i_1=1}^{L} \rho_{i_1} ( E_{21} ) Q \bigl( ~~
\sum_{j_1 =1}^{i_1-1}   \rho_{j_1} (-  {  H_1 -  H_2 \ovt }) ~+~ 
\sum_{j_1 = i_1 +1}^{ L }  \rho_{j_1} (  {  H_1 -  H_2  \ovt }) ~~ \bigr) \cr 
& \sum_{i_2=1}^{L} \rho_{i_2} ( E_{21} ) Q \bigl( ~~ 
\sum_{j_2 =1}^{i_2-1}   \rho_{j_2} (-  {  H_1 -  H_2  \ovt }) ~+~ 
     \sum_{j_2 = i_2 + 1}^{ L }  \rho_{j_2} (  {  H_1 -  H_2  \ovt } ~~ \bigr) 
) \cr 
&  \qquad\qquad \qquad\qquad\qquad
\cdots\cdots\cdots\cdots \cdots\cdots\cdots\cdots \cr 
&  \qquad\qquad \qquad\qquad\qquad
\cdots\cdots\cdots\cdots \cdots\cdots\cdots\cdots \cr 
& \sum_{i_n=1}^{L} \rho_{i_n} ( E_{21} ) Q \bigl(~~  
\sum_{j_n =1}^{i_n-1}   \rho_{j_n} (-  {  H_1 -  H_2  \ovt })
  ~ + ~\sum_{j_n = i_n +1}^{ L }  \rho_{j_2} (  {  H_1 -  H_2  \ovt } 
~~ \bigr)  \cr 
}} 
where for clarity we have used a definition $ Q( \rho_j ( H )   ) \equiv  
q^{ \rho_j ( H ) } $. 
It is convenient to bring all the $H$ factors to the right. 
In so doing we have to compute some commutators. 
The commutator terms coming from the $j_1$ sum 
are non-trivial when the 
$j_1$ is equal to $i_2$ or $ i_3$ up to $i_n$. 
The commutator terms from $ j_2$ are non-trivial
when $j_2$ is equal to $i_3, i_4 \cdots i_n$. 
Let us focus on the terms we get when 
$j_1$ is equal to $i_2$. 
Using $q^{H_1-H_2\ovt } E_{21} = q^{-1}  E_{21}q^{H_1-H_2 \ovt } $
we find that these commutator terms are
\eqn\comtmsot{ 
Q ~~ \bigl( ~~ \sum_{j_1 = 1}^{i_1-1} \delta ( j_1, i_2 ) - 
\sum_{j_1 = i_1 + 1}^{ L } \delta ( j_1, i_2 ) ~~ \bigr)  } 
If we define $ \theta_+ ( x) = 1 $ for integers $x \ge 1$ 
and zero otherwise, then 
we can write the above as 
\eqn\thp{ 
Q ~~\bigl( ~~   \theta_+ ( i_1 -  i_2 ) - \theta_+ (  i_2 - i_1 )
~~ \bigr ) 
= Q ~~ \bigl (  \theta  ( i_1 -  i_2 ) ~~ \bigr )   } 
We have also defined $ \theta ( x ) \equiv \theta_+ ( x) - \theta_+ ( - x)
$.  

Now we can write 
\eqn\simpd{\eqalign{  
& \Delta_q ( E_{21}^n ) = 
\sum_{i_1, i_2 \cdots i_n = 1 }^{L} \rho_{i_1} ( E_{21} ) \cdots
 \rho_{i_n} ( E_{21} ) \cr 
&  Q \bigl( ~~
\sum_{j_1 =1}^{i_1-1}   \rho_{j_1} (-  {  H_1 -  H_2 \ovt })
  ~+~ 
  \sum_{j_1 = i_1 + 1}^{ L }  \rho_{j_1} (  {  H_1 -  H_2  \ovt }) ~~
\bigr)
\cr 
& \qquad\qquad\qquad\qquad\qquad\quad ~ \vdots \cr 
&   Q \bigl( ~~
\sum_{j_n =1}^{i_n-1}   \rho_{j_n} (-  {  H_1 -  H_2 \ovt })
~ + ~\sum_{j_n = i_n +1}^{ L }  \rho_{j_2} (  {  H_1 -  H_2  \ovt } 
~~ \bigr) \bigr)\cr 
& Q \bigl( ~~ \sum_{ 1 \le k < l \le n } \theta (  i_k - i_l ) 
~~ \bigr )  \cr 
}}

When we act on $ \Phi_1^{ L} $ the $q^H$ factors which have 
been commuted through to the right are easily 
evaluated to give $ q^{{ ( L+1)n \ovt} - ( i_1 + \cdots i_n )  } $.  
In the sums above we have a restriction $ i_1 \ne i_2 \cdots \ne i_n $. 
This follows because $E_{21}^2 \Phi_1 = 0 $. 
The sum includes all possible orderings of the 
$i_1 \cdots i_n $ which can be described using 
permutations $\sigma $ in $S_n$. 
\eqn\intprm{ 
\sum_{i_1 \ne i_2 \cdots \ne i_n } 
~~~ = ~~~ \sum_{ \sigma} ~~~~~~  \sum_{ i_{\sigma(1)}  < i_{\sigma(2)}  \cdots
< i_{\sigma(n)} } } From \intprm\ we can write \simpd\ as 
\eqn\simpdi{\eqalign{  
& \Delta_q ( E_{21}^n ) ~ \Phi_1^{ \otimes L }
= \sum_{ \sigma }~~~ \sum_{ i_{\sigma(1)}   < i_{\sigma(2)}  \cdots
< i_{\sigma(n) } }
 q^ { { ( L +1) n \ovt } - ( i_1 + i_2 + \cdots i_n ) }  \cr 
& \qquad \qquad \qquad \qquad \qquad \qquad 
 \rho_{i_1} ( E_{21} ) \cdots \rho_{i_n} ( E_{21} ) ~ \Phi_1^{ \otimes L } \cr
& \qquad \qquad \qquad \qquad \qquad Q ~~\bigl ( 
 \sum_{ 1 \le k < l \le n }  
\theta ( ~ \sigma^{-1} ( k ) - \sigma^{-1} ( l ) ~ ) ~~\bigr ) \cr 
& = \sum_{ \sigma }~~~ Q ~~\bigl ( 
 \sum_{ 1 \le k < l \le n }   \theta ( ~ \sigma^{-1} ( k ) -
\sigma^{-1} ( l ) ~ ) ~~\bigr )
\sum_{ i_{\sigma(1)}   < i_{\sigma(2)}  \cdots
< i_{\sigma(n) } }
 q^ { { (L +1) n \ovt } - ( i_{\sigma(1)} + i_{\sigma(2)} + 
\cdots i_{\sigma(n) } )  }  \cr 
& \qquad \qquad \qquad \qquad \qquad \qquad 
 \rho_{i_{\sigma(1) }} ( E_{21} ) \cdots \rho_{i_{\sigma(n) } }
( E_{21} )
~ \Phi_1^{ \otimes L } \cr
& \qquad \qquad \qquad \qquad \qquad \cr 
&  = \sum_{ \sigma }~~~  Q ~~\bigl ( 
 \sum_{ 1 \le k < l \le n }  
 \theta ( ~ \sigma^{-1} ( k ) - \sigma^{-1} ( l ) ~ ) ~~\bigr )
\sum_{ i_1   < i_2  \cdots
< i_n }
 q^ { { (L +1) n \ovt } - ( i_1 + i_2 + 
\cdots i_n)  }  \cr 
& \qquad \qquad \qquad \qquad \qquad \qquad 
 \rho_{i_1 } ( E_{21} ) \cdots \rho_{i_n } ( E_{21} ) ~ \Phi_1^
{ \otimes L }
\cr 
& = [n]! \sum_{ 1 \le i_1  < i_2 \cdots < i_n  \le   L  } 
 q^ { { (L +1) n \ovt } - ( i_1 + i_2 + 
\cdots i_n)  }  
 \rho_{i_1 } ( E_{21} ) \cdots \rho_{i_n } ( E_{21} ) ~ \Phi_1^
{ \otimes L }
\cr 
}}
In the second equality we have recognized that the 
phase factor coming from the commutations only depends 
on the ordering on the $i$'s and hence only on the permutations 
$ \sigma$, so they can be factored out of the $i$ sum. 
In the next equality, we use the symmetry of the 
summand of the sum over $i$ in order to replace $i$ with 
$i_{\sigma}$. In the next line we renamed the summation variable. 
Finally the sum over  $\sigma$ was evaluated to 
give a constant $q$-factorial $[n]_q!$ which is defined 
as $ [n]_q! = [n]_q [n-1]_q \cdots [2]_q [1]_q $ where $[k]_q = { q^{k} - q^{-k}
\over q - q^{-1} }$. 
Note that it is invariant under $q \rightarrow q^{-1}$
which is as it should be since, in the sum, changing $q$ to $q^{-1}$ 
is equivalent to exchanging $\sigma $ with $ \sigma^{-1}$. 

With this form of the $\Delta_q (E_{21}^n ~ )~~
  \Phi_1^{ \otimes L } $ we have a sequence of 
operators in tensor space. We multiply 
them and take a trace. Denoting the 
 combined operation as $ TR $ we find 
\eqn\tdelp{\eqalign{  
& TR ~(~  ~ \Delta_q (E_{21}^n ~ )~~
  \Phi_1^{ \otimes L }   ) \cr   
& =  q^ { (L+1) n \ovt }[n]_q!~~ \sum_{1 \le i_1  < i_2 \cdots < i_n  \le   L } 
q^ { - ( i_1 + i_2 + \cdots i_n)  }~~~ TR  ~~\bigl (~~ \rho_{i_1 } ( E_{21} )
\cdots  \rho_{i_n } ( E_{21} )  ~~~ \Phi_1^{\otimes L } ~~\bigr ) ~~  \cr 
&=  q^ { J n \ovt }[n]_q!~~ \sum_{1 \le j_1  \le  j_2 \cdots \le 
 j_n  \le   J  } q^ { - ( j_1 + j_2 + \cdots j_n)  } 
TR  [j_1, j_2-j_1, j_3-j_2, \cdots , j_n -j_{n-1} , J - j_n ] 
\cr }}
Here we have defined $ j_l = i_l -l $ for $l=1, \cdots, n $ 
and have used the notation in \opnoti\ in the last line. 
The upper limit of $j_n$ is now $ J = L -n $. 
We now introduce variables $ k_1 = j_{2 } - j_{1} $
and $k_l - k_{l-1} = j_{l+1 } - j_{l} $ for $  2 \le l \le n-1 $.   
We can rewrite the previous formula as 
\eqn\rewrt{\eqalign{   
& q^ { J n \ovt }[n]_q!~~ \sum_{1 \le k_1 \le k_2  \cdots \le 
 k_{n-1}  \le   J  } 
\sum_{ j_1 = 0 }^{J-k_{n-1} } 
 q^{-nj_1} q^ { - (  k_1 + k_2  + \cdots k_{n-1} )  } \cr 
& \qquad \qquad 
TR (  [, k_1, k_2- k_1 , \cdots , k_{n-1}  -k_{n-2} , J - k_{n-1}  ]
  ) \cr }}
where the upper limit on the $j_1$ sum 
is easily fixed by requiring that the sums 
in \rewrt\ and \tdelp\ have the same number of terms. 
After doing the sum over $j_1$  we get 
\eqn\tdelpres{\eqalign{  
& TR ~(~  ~ \Delta_q (E_{21}^n ~ )~~
  \Phi_1^{ \otimes L }   ) \cr 
& =   q^ { J n \ovt }[n]_q!  
\sum_{1 \le k_1 \le k_2  \cdots \le 
 k_{n-1}  \le   J  } { ( 1 - q^{ -n( 1 - k_{n-1} ) } ) \over ( 1 -
  q^{-n} )} 
 q^ { - (  k_1 + k_2  + \cdots k_{n-1} )  }  \cr 
& \qquad \qquad TR (  [, k_1, k_2- k_1 , \cdots , k_{n-1}  -k_{n-2} , J - k_{n-1}  ] )
  \cr }} 
The term involving $q^{ -n( 1 -k_{n-1} )} $  does not look symmetric 
but actually is symmetric after we use cyclicity.  We show this
in more detail for the two $\tau$ case later. 
The BMN operator corresponding to  $TR ~(~  ~ \Delta_q (E_{21}^n ~ )~~
  \Phi_1^{ \otimes L }   )$ is therefore 
$ (-1)^{  n }[n]_q !  \alpha^{\dagger}_{n-1} 
( \alpha^{\dagger}_{-1})^{n-1}  | 0 > $, where the 
denominator $( 1 - q^{-n} ) $ has cancelled after we combined 
contributions from the two terms in the numerator, and 
we used $ q^{nJ \ovt  } = (-1)^n $. 

\subsec{ The case of a single $\tau$ }
 
We now move on to consider the insertion of a single
$\tau$ operator of the type defined in section 4, i.e., we want to compute
$ TR  \tau_{a,0 } \Delta_q (E_{21}^n )$ 
The first step is to consider
the operator $ \tau_{a,0 } \Delta_q (E_{21}^n )$,
which is an element 
of $ \cA^{\otimes L } $ where $\cA$ is the algebra of Higgs fields. 
To evaluate the trace we apply the multiplication map to 
get an element of $ \cA $ from the element in $ \cA^{\otimes L } $. 
Then we take a trace of this element. As we saw in \formqcyc, 
the $q$-cyclic $\tau $ operator can be written as a 
sum of cycling operations weighted by $q$ factors 
which depend on the $U_q(2)$ quantum numbers of the 
elements cycled. We are now composing $ \tau_{a,0 }$ from 
the left with  $\Delta_q (E_{21}^n )$. It is useful to 
keep the cycling operators on the left but to commute 
the $H$-factors to the right. Since we are calculating 
a trace at the end, cycling operations on the left can be set to $1$. 
On the other hand since we are acting on $ \Phi_1^L $ on the right 
the $H$ factors are easy to evaluate. With this in mind and
using \simpdi\ and \formqcyc\ we 
expand : 
\eqn\taudel{\eqalign{  
& \tau_{a,0 } \Delta_q (E_{21}^n ) \Phi_{1}^{\otimes L} \cr 
& = [n]_q!~ q^ {  (L +1) n \ovt }~ \sum_{k=1}^{L}  ~ c^k ~   Q ~~ 
\bigl( ~~  \sum_{l=1}^{k} ~~ \rho_{l} ~(~  aH_1  ~) ~~ \bigr )~~ \cr 
& \sum_{1 \le i_1  < i_2 \cdots < i_n  \le   L }  q^ {   - ( i_1 + i_2 + 
\cdots i_n)  } ~~ \rho_{i_1 } ( E_{21} ) \cdots 
 \rho_{i_n } ( E_{21} )  \Phi_{1}^{\otimes L}            \cr 
&= [n]_q!~ q^ {  ( L +1) n \ovt }
\sum_{k=1}^{L} c^k \sum_{1 \le i_1  < i_2 \cdots < i_n  \le   L } 
 q^ { - ( i_1 + i_2 + \cdots i_n)  } ~~~ \rho_{i_1 } ( E_{21} )
\cdots  \rho_{i_n } ( E_{21} )   \cr 
&  \qquad \qquad \qquad 
Q ~~ \bigl( ~~ \sum_{l=1}^{k} \rho_{l} ~(~ aH_1  ~)~ -  a ~ \sum_{l=1}^{k} 
 ~~( ~~ \delta (l,i_1 ) + \cdots \delta
( l, i_n ) ~~)  ~~~\bigr ) ~ \Phi_{1}^{\otimes L} \cr 
&=[n]_q! ~q^ {  (L+1) n \ovt }
\sum_{k=1}^{L} c^k \sum_{1 \le i_1  < i_2 \cdots < i_n  \le   L } 
 q^ { - ( i_1 + i_2 + \cdots i_n)  }\rho_{i_1 } ( E_{21} ) 
 \cdots  \rho_{i_n } ( E_{21} )  \cr 
&\qquad \qquad \qquad 
 Q ~~ \bigl( ~~   ak   - ~a~ 
 \sum_{l=1}^{n}   \theta_+ ( k - i_l + 1 ) ~~
\bigr )~ \Phi_{1}^{\otimes L}  }}
Acting with the trace 
we get  
\eqn\ttaudelp{\eqalign{  
& TR ~(~  \tau_{a,0 } ~ \Delta_q (E_{21}^n ~ )~~
  \Phi_1^{ \otimes L }   ) \cr   
& =  q^ { (L+1) n \ovt }[n]_q!~~ \sum_{1 \le i_1  < i_2 \cdots < i_n  \le   L } 
q^ { - ( i_1 + i_2 + \cdots i_n)  }~~~ TR  ~~\bigl (~~ \rho_{i_1 } ( E_{21} )
\cdots  \rho_{i_n } ( E_{21} )  ~~~ \Phi_1^{\otimes L } ~~\bigr ) ~~  \cr 
&\qquad \qquad \sum_{k=1}^{L} ~~~ q^{ a k } ~~
  Q ~~ \bigl( ~~   - a ~(~  \theta_+ ( k - i_1 + 1 ) + \cdots 
\theta_+ ( k - i_n + 1 ) ~)~   ~~
\bigr )~~  \cr 
}}
The sum over $k$ can be written out 
as 
\eqn\theksum{\eqalign{  
& \sum_{k=1}^{L} q^{ a k } ~~ Q ~~ \bigl( ~~   - a ~ 
( ~ \theta_+ ( k - i_1 + 1
)~ +~ \cdots ~ + ~ \theta_+ ( k - i_n + 1 ) ~ )~  ~~ \bigr )  \cr 
& = \sum_{k=1}^{i_1-1} q^{ a k } ~+~  q^{-a} \sum_{k=i_1}^{i_2-1 }
   q^{ a k }
   ~+~  q^{-2a} \sum_{k=i_2}^{i_3-1}  q^{ a k } ~+~  \cdots \cr  
&    \qquad \qquad  + q^{-(n-1)a} \sum_{k=i_{n-1} }^{i_{n}-1} q^{ a k }
   ~+~  q^{-na}  \sum_{k= i_n }^{L}  q^{ a k } \cr 
& = { q^a ( 1 - q^{a(L-n)} ) \over ( 1 - q^a ) }   ~+~  ( q^{a ( i_1-1 ) } 
   ~+~  q^{ a ( i_2 -2 ) } ~+~ ~  \cdots ~~ +~  q^{ a ( i_n - n  ) } )  \cr 
& =  { q^a ( 1 - q^{a(L-n)} ) \over ( 1 - q^a ) } ~~+~~  
     \sum_{l=1}^{n } q^{a j_l }  \cr }} 
In the last line we have used $ j_l \equiv i_l - l $. 
When we are constructing BMN operators we 
use $ q^{L-n} = q^{J} = 1 $, which means that the constant term is 
zero.

 We can now write a simpler expression for the 
 result of acting on $ \Phi_1^{ \otimes L } $ 
 with the quantum group generators, the q-cyclic 
 operator and the trace  : 
\eqn\simpttdp{\eqalign{  
& TR ~(~  \tau_{a,0 } ~ \Delta_q (E_{21}^n ~ )~~
  \Phi_1^{ \otimes L }   ) \cr 
& = [n]_q!~ q^ {  J n \ovt } 
 \sum_{ 0  \le j_1  \le  j_2 \cdots \le  j_n  \le   J   } 
 \bigl ( \sum_{l=1}^{n } q^{a j_l } \bigr ) 
q^ { - ( j_1 + j_2 + \cdots j_n)  } ~~~ \cr 
&  ~~ TR  ~~\bigl (~~  \rho_{j_1+1 } ( E_{21} ) \cdots  \rho_{j_n + n  }
 ( E_{21} ) ~~ 
\Phi_1^{\otimes L }  ~~\bigr ) ~~ \cr 
& = [n]_q!~ (-1)^ {  n } 
 \sum_{ 0  \le j_1  \le  j_2 \cdots \le  j_n  \le   J   } 
 \bigl ( \sum_{l=1}^{n } q^{a j_l } \bigr ) 
q^ { - ( j_1 + j_2 + \cdots j_n)  } ~~~ \cr 
&  ~~ TR  ~~ [ ~ j_1 , j_2 -j_1, \cdots, j_n -j_{n-1} , J -j_n ~ ]  ~~ 
\cr }}
In the final line we have used the notation of \opnoti. 
It is useful to define a new set of summation variables 
$ ( j_1, k_1 , k_2, \cdots , k_{n-1} )$ to replace 
the set $ (j_1, j_2, \cdots , j_n )$. They 
are defined as follows 
\eqn\defvar{\eqalign{  
&  k_1 =  j_2 - j_1  \cr 
&  k_2-k_1 = j_3- j_2 \cr 
& \vdots \cr  
& k_{n-1} - k_{n-2} = j_n - j_{n-1} \cr 
}}
which imply 
\eqn\defvar{\eqalign{  
&  k_1  =  j_2 - j_1 \cr 
&  k_2 =  j_3 -  j_1 \cr 
& \vdots \cr  
& k_{n-1}  =  j_n -  j_1 \cr 
}}
Now the sum can be manipulated to 
\eqn\summanip{\eqalign{  
& \sum_{  j_1=0}^{ J } \sum_{  j_2 = j_1 }^{J } \sum_{j_3= j_2}^{J } 
 \cdots \sum_{j_n=j_{n-1}}^{J } \cr 
& = \sum_{k_1=0}^{J} \sum_{k_2=k_1}^{J} \cdots \sum_{k_{n-1}=k_{n-2}}^{J} 
   \sum_{j_1= 0}^{J- k_{n-1}} \cr }}
After doing the sum over $j_1$ we are left with 
\eqn\finfrmot{\eqalign{  
& TR ~(~  \tau_{a,0 } ~ \Delta_q (E_{21}^n ~ )~~
  \Phi_1^{ \otimes L }   ) \cr 
& = { [n]_q! q^{nJ\over 2 } 
\over ( 1 - q^{a-n} ) } \sum_{0 \le k_1 \le k_2 \cdots \le k_{n-1} \le J } 
   { ( 1 - q^{ (a -n) (  + 1 - k_{n-1} )  } )  } ( 1 + q^{a k_1} + 
q^{a k_2} + \cdots + q^{a k_{n-1} } ) \cr 
&     q^{-k_1 - k_2 - \cdots - k_{n-1} } ~
TR [,k_1,k_2 - k_1, ...,k_{n-1} - k_{n-2}, J-k_{n-1}]
 }} 
where we have written the trace of the operator
piece of the expression in the notation introduced
in section 2.
When we look at the term 
$ ( 1 + q^{ak_1} + 
q^{ak_2} + \cdots + q^{a k_{n-1} } ) $ it is clearly 
symmetric under permutations of $k_1 $ to $k_{n-1}$. 
It corresponds to the string state
$ \alpha_{n-1} \alpha_{-1}^{n-2}
  + \alpha_{n-1+a}   \alpha_{-1 +a } \alpha_{-1}^{n-2} $
by the BMN map. 
The term   
\eqn\qterm{\eqalign{  
& \sum_{0 \le k_1 \le k_2 \cdots \le k_{n-1} \le J }
q^{ (a -n)  } q^{ - ( a - n )  k_{n-1} } ( 1 + q^{ak_1} + 
q^{ak_2} + \cdots + q^{a k_{n-1} } )   q^{-k_1 - \cdots - k_{n-1} }    \cr 
& \qquad \qquad \qquad \qquad 
TR [,k_1,k_2 - k_1, ...,k_{n-1} - k_{n-2}, J-k_{n-1}]
}}
does not appear manifestly symmetric, 
but we can, by using cyclicity, write it 
as 
\eqn\qtermi{\eqalign{ 
\sum_{0 \le k_1 \le k_2  \cdots \le k_{n-1} \le J } &
{ q^{ (a -n)  }  \over (n-1)  }  q^{-k_1 - \cdots - k_{n-1} }
( q^{-(  a-n )k_{n-1} } + 
q^{-(  a-n )k_{n-2}  } + \cdots + q^{ -(a-n)k_1 } )  \cr
& ( 1 + q^{ak_1}  + \cdots + q^{a k_{n-1} } )  ~~ 
TR [,k_1,k_2 - k_1, ...,k_{n-1} - k_{n-2}, J-k_{n-1}]
}}
In the next section we elaborate on the cycling 
manipulations in the two $\tau$ case. 
Now the result is  obviously symmetric 
and the entire operator corresponds to a sum of BMN states of the 
form $ \alpha_{n-1} \alpha_{-1}^{n-2} + 
\alpha_{n-1+a}   \alpha_{-1 +a } \alpha_{-1}^{n-2} | 0 \rangle $.

\subsec{ Two $\tau$-operators } 

We now consider the action of 
two $\tau$-operators on \simpdi, using the form 
\formqcyc\ for $\tau $ operators.  Explicitly we have 
\eqn\formttdel{\eqalign{  
& \tau_{a_2,0 } \tau_{a_1,0}  \Delta_q (E_{21}^n ) \Phi_{1}^{\otimes L}  \cr 
& = [n]_q!~ q^ {  (L +1) n \ovt }~ \sum_{k_2=1}^{L}  ~ c^{k_2} ~   Q ~
\bigl( ~  \sum_{l_2=1}^{k_2} ~~ \rho_{l_2} ~(~  a_2 H_1  ~) ~ \bigr )~
\sum_{k_1=1}^{L}  ~ c^{k_1} ~   Q ~~ 
\bigl( ~~  \sum_{l_1=1}^{k_1} ~~ \rho_{l_1} ~(~  a_1 H_1  ~) ~~ \bigr )~~
\cr 
& \sum_{1 \le i_1  < i_2 \cdots < i_n  \le   L } 
 q^ {   - ( i_1 + i_2 + 
\cdots i_n)  } ~~ \rho_{i_1 } ( E_{21} ) \cdots 
 \rho_{i_n } ( E_{21} )  ~~~  \Phi_{1}^{\otimes L}      \cr 
& =  [n]_q!~ q^ {  (L +1) n \ovt }~ \sum_{k_1,k_2 =1}^{L}  ~ c^{k_1+k_2} ~
\sum_{1 \le i_1  < i_2 \cdots < i_n  \le   L } 
 q^ {   - ( i_1 + i_2 + 
\cdots i_n)  } ~~ \rho_{i_1 } ( E_{21} ) \cdots 
 \rho_{i_n } ( E_{21} )  \cr  
&   
Q ~ \bigl( ~~ \sum_{l_1=1}^{k_1} \rho_{l_1} ~(~ a_1H_1  ~)~ - 
 \sum_{l_1=1}^{k} 
 ~~( ~~ \delta (l_1,i_1 ) + \cdots + \delta
( l_1, i_n ) ~~)  ~~\bigr )  \cr 
&    
Q ~ \bigl(~~ \sum_{l_2=1}^{k_2} \rho_{r_{L} ( l_2+ k_1 )  }
 ~(~ a_2H_1  ~)~ -  
\sum_{l_2 = 1}^{k_2} 
 ~( ~ \delta ( r_L ( k_1 + l_2 ) ,i_1 ) + \cdots + \delta
( r_L ( k_1+ l_2 ) , i_n ) ~) ~\bigr ) ~ \Phi_{1}^{\otimes L}  \cr 
&=[n]_q! ~q^ {  (L+1) n \ovt }
\sum_{k_1,k_2=1}^{L} 
 c^{k_1+k_2}  \sum_{1 \le i_1  < i_2 \cdots < i_n  \le   L } 
 q^{ - ( i_1 + i_2 + \cdots i_n)  }\rho_{i_1 } ( E_{21} ) 
 \cdots  \rho_{i_n } ( E_{21} )  \cr 
& \quad 
 Q ~~ \bigl( ~~   \sum_{l_1=1}^{k_1} \rho_{l_1} ( a_1H_1  )  - ~a_1~ 
 \cF ( k_1, 0 ; \vec i ) ~~~ 
\bigr )~~ \cr
& \quad  
 Q ~~ \bigl( ~~   \sum_{l_2=1}^{k_2} \rho_{ r_L ( l_2 + k_1) } ( aH_1  )  - 
~a_2~  \cF (k_2, k_1 ; \vec i )~~~ 
\bigr )~ \Phi_{1}^{\otimes L} 
}}
The manipulations are similar to those 
in \taudel.  One new thing is that 
when the $ \rho_{l_2} ( a_2 H_1) $ is commuted past the 
$c^{k_2}$ it becomes  $ \rho_{l_2+k_1} ( a_2 H_1) $ 
if $ 1 \le l_2 + k_1 \le L  $  but $ \rho_{l_2+k_1- L } ( a_2 H_1) $
if $  l_2 + k_1 >    L $. This has been  conveniently written as 
$   \rho_{ r_L ( l_2+k_1 )  } ( a_2 H_1) $ where $ r_L ( m ) $ 
 for an integer $m$ is defined as one plus the residue modulo $L$
of $m$.   
There is a consequent change in the sum over 
delta's. We have introduced functions 
$ \cF (p,q ; \vec i ) $ which depend on two positive integers 
$p,q$ and a fixed set of integers $i_1, \cdots, i_n $ 
between $ 1$ and $L$. 
It is defined by
\eqn\defF{ 
\cF  ( p,q ~;~  \vec i ) = \sum_{l=1}^{n} 
\sum_{m=q+1}^{q+p } \delta ~(~  i_{l}, r_L ( m ) ~ ) .} 
The function $\cF ( p,q ~;~ \vec i )  $  counts the number of $i_l$'s 
satisfying $ r_L (q + 1) \leq i_l \leq r_L (q+p)$. 
The sums of delta's 
are naturally written in terms of these 
functions. The functions $\cF $  satisfy a useful property 
\eqn\usful{ 
\cF  (p+q, 0 ~;~    \vec i ) = \cF (p , 0 ~;~ \vec i ) + 
\cF (q, p ~;~ \vec i ) 
} 

In \formttdel\ 
the $H$ factors on the right 
are easily evaluated to give $q^{a_2 k_2 + a_1 k_1} $, moreover,
taking the trace allows
the $c^{k_1 +k_2} $ to be set to one due to cyclicity. 
\eqn\ttdelt{\eqalign{ 
& TR ~(~  \tau_{a_2, 0 } \tau_{a_1,0 } ~ \Delta_q (E_{21}^n ~ )~~
  \Phi_1^{ \otimes L }   ) \cr   
& =  q^ { (L+1) n \ovt }[n]_q!~~ \sum_{1 \le i_1  < i_2 \cdots < i_n  \le   L } 
q^ { - ( i_1 + i_2 + \cdots i_n)  }~~~ TR  ~~\bigl (~~ \rho_{i_1 } ( E_{21} )
\cdots  \rho_{i_n } ( E_{21} )  ~~~ \Phi_1^{\otimes L } ~~\bigr ) ~~  \cr 
&\qquad \qquad \sum_{k_1,k_2=1}^{L} ~~~ q^{ a_1 k_1 +a_2 k_2 } ~~
 q^{ - a_1 \cF (k_1, 0 ~;~ \vec i ) - a_2 \cF  ( k_2, k_1 ~;~ \vec i ) } \cr 
}}
Performing the sum over $k_2$ 
 we find 
 \eqn\ktsm{ 
 \sum_{k_2 = 1 }^{L} q^{a_2 k_2 - a_2 \cF ( k_2, k_1 ~;~ \vec i ) } 
 =  q^{  -a_2 ~(~ k_1 - \cF (k_1, 0 ~;~ \vec i) ~ ) } 
  ~~ \cS ( a_2 ~;~  \vec i  ) } 
where $\cS(a  ~;~ \vec i ) $ is defined to be the sum 
\eqn\defS{ { \cal {S}} ( a ~;~  \vec i ) = 
\sum_{l=1 }^{n} q^{ a ( i_l -l ) }. }
 While the summation index $k_2$ in \ktsm\ is
 constrained by $1 \leq k_2 \leq L$, the integer 
 $k_1$ can be outside this range.  The same basic
sums will be used over and over again as we increase
the number of $\tau$'s in the next section.
They are evaluated by similar methods 
to those used in the previous subsection, taking advantage of $q^{ J} = 1$. 

The $k_1$ sum now follows as a special case of \ktsm\ and is given by 
\eqn\kosm{ 
\sum_{k_1 = 1 }^{L} q^{ ( a_1 -a_2 )  ~(~ 
 k_1 -  \cF  ( k_1, 0 ~;~ \vec i ) ~)~ } 
= \cS ( a_1 -a_2 ~;~ \vec i   ). 
} 
The result is therefore 
\eqn\resttd{\eqalign{ 
& TR ~(~  \tau_{a_2, 0 } \tau_{a_1,0 } ~ \Delta_q (E_{21}^n ~ )~~
  \Phi_1^{ \otimes L }   ) \cr   
& =  q^ { (L+1) n \ovt }[n]_q!~~ \sum_{1 \le i_1  < i_2 \cdots < i_n  \le   L } 
q^ { - ( i_1 + i_2 + \cdots i_n)  }~~ 
\cS ( a_1 -a_2 ~;~ \vec i  ) \cS ( a_2 ~;~ \vec i ) \cr
& TR [i_1 -1, i_2 - i_1 -1,...,i_n - i_{n-1} - 1, L - i_n] \cr  
& =  q^ { J  n \ovt }[n]_q!~~ \sum_{1 \le j_1  \le j_2 \cdots \le j_n  \le   J } 
q^ { - ( j_1 + j_2 + \cdots j_n)  }~~ 
\cS ( a_1 -a_2 ~;~  j_1 +  1, \cdots , j_n + n    ) \cr 
& \qquad \qquad \qquad  \cS ( a_2 ~;~ j_1 + 1 , \cdots j_n +n  )  ~~ 
 TR [j_1 , j_2 - j_1,...,j_n - j_{n-1} , J  - j_n] \cr  }}

In the last line we changed variables $j_l  = i_l -l $.  
This resulting  expression is not quite of 
the BMN form given in \BMNtwo, however we can make the same
basic manipulations described in the formulae \defvar\ and
\summanip\ to reach such a form.  First we use cyclicity of the trace
to move the $j_1$ powers of $\Phi_1$ to the right,
 and then redefine summation indices
as $k_1 = j_2 - j_1 $ and $k_l - k_{l-1} = j_{l+1} - j_l $ for
$2 \leq l \leq n-1$.  Equivalently we find
$j_{l+1} =  k_l + j_1$ for $1 \leq l \leq n-1$.
The operator \resttd\ becomes
\eqn\resttdtwo{\eqalign{ 
& TR ~(~  \tau_{a_2, 0 } \tau_{a_1,0 } ~ \Delta_q (E_{21}^n ~ )~~
  \Phi_1^{ \otimes L }   ) \cr   
& =  q^ {Jn/2} [n]_q!~~ 
\sum_{0 \le k_1  \le k_2 \cdots \le
k_{n-1}  \le   J } \sum_{j_1 =0 }^{J - k_{n-1} } q^{(a_1 - n)(j_1)}
q^ { - ( k_1 + k_2 + \cdots k_{n-1} )  }~~ \cr
& \cT ( a_1 -a_2 ~;~ \vec k  ) \cT ( a_2 ~;~ \vec k ) ~
TR [, k_1, k_2 - k_1,...,k_{n-1} - k_{n-2}, J - k_{n-1}] }}
where we have introduced the function ${\cal T}(a; {\vec k})$
defined as
\eqn\defnT{{\cal T}(a; {\vec k}) = 1 + \sum_{l=1}^{n-1} q^{a k_l}}
and which is related to $\cS$ by
\eqn\STreln{\cS (a; j_1+ 1 , \cdots , j_n + n ) =
 q^{a j_1 } \cT (a; {\vec k})}
given the change of variables above.
The $j_1$ sum can now be done trivially and is proportional
to $1 - q^{(a_1 - n)(1 - k_{n-1})}$.  Up to the factor
of $\cT (a_1 - a_2; \vec k)$,
this is exactly
what we found in the single $\tau$ case in the previous
section in \finfrmot\ if one identifies $a$ with
$a_2$.  Exactly as in that case, we find that the 
coefficient of the operator $TR [,k_1,...,J-k_{n-1}]$
is symmetric under permutations of the $k_l$'s
except for the factor of $q^{(a_1 - n)(1 - k_{n-1})}$
which arises from the $i_1$ sum.  This term nevertheless
can be made symmetric by using cyclicity of the trace
and redefining summation indices.
To see this, cyclicity allows us to rewrite the
operator as
\eqn\rewrite{TR [,k_1,k_2 - k_1,...,J-k_{n-1}]
= TR [,k_2 - k_1, k_3 - k_2, ..., J-k_{n-1},k_1].}
To put this operator back into the standard form
we define a new set of summation indices
\eqn\ktilde{\eqalign{
\kt_1 & = k_2 - k_1 \cr
\kt_l - \kt_{l-1} & = k_{l+1} - k_l, ~ 2 \leq l \leq n-2 \cr
\kt_{n-1} - \kt_{n-2} & = J - k_{n-1},}}
or equivalently solving for the $\kt$ variables
\eqn\solvekt{\eqalign{
\kt_l & = k_{l+1} - k_1 , ~ 1 \leq l \leq n-2 \cr
\kt_{n-1} & = J - k_1 .}}
Under this change of variables the various $q$-factors
transform as
\eqn\transf{\eqalign{q^{-(k_1 + \cdots + k_{n-1})} & = q^{n \kt_{n-1}}
q^{-(\kt_1 + \cdots + \kt_{n-1})} \cr
{\cal T}(a; {\vec k(\kt)}) & = q^{ - a \kt_{n-1}}{\cal T}(a; {\vec \kt})
\cr
q^{(a_1 - n)(1 - k_{n-1})} & =q^{(a_1 - n)(1 - \kt_{n-2} +
\kt_{n-1})}}}
so that the  $q$-factor  transforms as
\eqn\wholetransf{\eqalign{ q^{(a_1 - n)(1 - k_{n-1})} &
q^{-(k_1 + \cdots + k_{n-1})}
\cT ( a_1 -a_2 ~;~ \vec k  ) ~ \cT ( a_2 ~;~ \vec k ) \cr
& = q^{(a_1 - n)(1 - \kt_{n-2})} q^{-(\kt_1 + \cdots + \kt_{n-1})}
\cT ( a_1 -a_2 ~;~ {\vec \kt}  ) ~ \cT ( a_2 ~;~ {\vec \kt} ).}}
In the end the only effect of these operations is to change
the $k_{n-1}$ in the $q$-factor to $k_{n-2}$.
  Since these two different forms are
  equal, it means that this term  is actually  symmetric 
  under exchange of $k_{n-1} $ and $k_{n-2}$. 
Repeating this prodecure $n-3$ more times, we can rewrite
\resttdtwo\ in a manifestly symmetric form
\eqn\fintwotau{ \eqalign{ 
& TR ~(~  \tau_{a_2, 0 } \tau_{a_1,0 } ~ \Delta_q (E_{21}^n ~ )~~
  \Phi_1^{ \otimes L }   ) \cr   
& =  q^ {Jn/2} {1 \over 1 - q^{a_1 - n}} [n]_q!~~ 
\sum_{0 \le k_1  \le k_2 \cdots \le
k_{n-1}  \le   J } \left( 1 - {1 \over n-1} \sum_{l=1}^{n-1} 
q^{(a_1 - n)(1-k_l)}
\right) \cr
& q^ { - ( k_1 + k_2 + \cdots k_{n-1} )  }~
\cT ( a_1 -a_2 ~;~ \vec k  ) ~ \cT ( a_2 ~;~ \vec k ) ~
TR [, k_1, k_2 - k_1,...,k_{n-1} - k_{n-2}, J - k_{n-1}]. }}

This is our final form for the two $\tau$ operator.  Comparing to
the BMN operator \BMNtwo\ we see that \fintwotau\ consists
of a sum of BMN operators, specifically it consists of 
\eqn\twotauequiv{\eqalign{TR ~(~  \tau_{a_2, 0 } \tau_{a_1,0 } ~ 
\Delta_q (E_{21}^n ~ )~~
  \Phi_1^{ \otimes L }   ) & \leftrightarrow 
\alpha^{\dagger}_{n - a_1-1} \alpha^{\dagger}_{a_1 - a_2 -1} 
\alpha^{\dagger}_{a_2 -1} 
(\alpha^{\dagger}_{-1})^{n-3} \cr
& +  \alpha^{\dagger}_{n+ a_2 - a_1 -1} \alpha^{\dagger}_{a_1 - a_2 -1} 
 (\alpha^{\dagger}_{-1})^{n-2} \cr
& + \alpha^{\dagger}_{n - a_2 -1} \alpha^{\dagger}_{a_2 -1} 
 (\alpha^{\dagger}_{-1})^{n-2} \cr
& + \alpha^{\dagger}_{n - a_1 -1} \alpha^{\dagger}_{a_1 -1} 
 (\alpha^{\dagger}_{-1})^{n-2} \cr
& + \alpha^{\dagger}_{n-1} (\alpha^{\dagger}_{-1})^{n-1}}}
where the operator content on the right-hand-side is meant to
be schematic in that we have not tried to get the constants right.
Comparing to the single $\tau$ case the important point to note
is that there is now an operator on the RHS which has two generic
oscillator numbers.  In the single $\tau$ case the best one could
do was to get an operator with only one generic oscillator number.
We now show that this pattern continues - letting $P$ $\tau$'s
act on the operator \simpdi, we find a dual string state containing
in particular a state with $P$ generic oscillator numbers.
Consequently, acting with $n-1$ $\tau$'s will produce a dual
string state with completely generic oscillator numbers.

\subsec{ Three and more $ \tau$'s }

We now turn to the generic case of many $\tau$'s.
For simplicity we shall sketch the three $\tau$ case
and simply state the end result for the $n-1$ $\tau$ case.
The manipulations from the previous subsections are more
or less identical in these higher $\tau$ cases. 
Consider now the three $\tau$ case:
\eqn\thrta{\eqalign{  
&  TR ~(~ \tau_{a_3, 0 }  \tau_{a_2, 0 } \tau_{a_1,0 } ~
 \Delta_q (E_{21}^n ~ )~~
  \Phi_1^{ \otimes L }   ) \cr 
 & =  q^ { (L+1) n \ovt }[n]_q!~~ \sum_{1 \le i_1  < i_2 \cdots < i_n  \le   L } 
q^ { - ( i_1 + i_2 + \cdots i_n)  }~~~ TR  ~~\bigl (~~ \rho_{i_1 } ( E_{21} )
 \cdots  \rho_{i_n } ( E_{21} )  ~~~ \Phi_1^{\otimes L } ~~\bigr ) ~~  \cr 
&\qquad \qquad \sum_{k_1,k_2, k_3=1}^{L} ~~~ q^{ a_1 k_1 +a_2 k_2 + a_3k_3 } ~~
 q^{ - a_1  \cF (k_1, 0 ~;~ \vec i ) - a_2 \cF  ( k_2, k_1 ~;~ \vec i )    
     - a_3 \cF ( k_3, k_1 +k_2 ~;~ \vec i ) } \cr }} 
The $k_3$ sum gives 
\eqn\insm{\eqalign{  
&\sum_{k_3 = 1 }^{L} 
q^{a_3  ~(~  k_3 - \cF ( k_3, k_1 + k_2 ~; ~  \vec i ~) ~) }
\cr 
& = q^{ -a_3 ~( ~k_1 + k_2 -  \cF ( k_1+k_2, 0 ~;~  \vec i ~)~ ) }  
  \cS  ( a_3 ~;~  \vec i) \cr 
&= q^{ -a_3 ~(~ k_1 + k_2 -  \cF ( k_1, 0  ~;~  \vec i) - 
\cF (k_2, k_1~;~  \vec i  ~)~ ) }   
  \cS  ( a_3 ~; ~  \vec i) \cr 
 }} 
Note that the sum we have to do is of the same form 
as in the two-$\tau$ case \ktsm\ 
with $a_2$ in that equation replaced by $a_3$ 
and the $ \cF (k_2, k_1 ~;~ \vec i  )$    in \ktsm\ replaced by
 $ \cF (k_3, k_1 +k_2 ~;~ \vec i )$. 
In the last line we used the property 
of $\cF $ given in \usful. 
Now we do the sum over $k_2$ in \thrta\  
\eqn\secsm{\eqalign{  
&\sum_{k_2 = 1 }^{L}  q^{   ( a_2 -a_3) ( k_2 - 
\cF ( k_2, k_1 ~;~  \vec i )  ) }  \cr 
&   =  q^{   - (a_2-a_3) ( k_1  -  \cF ( k_1 , 0 ~;~  \vec i) ) } 
  \cS ( a_2 - a_3 ~;~ \vec i) \cr 
 }} 
Again this is the same sum as \ktsm\ 
with $a_2$  replaced by $a_2-a_3$. 
Collecting the $k_1$ dependent terms, the final sum over $k_1$ is
\eqn\smko{\eqalign{ 
&\sum_{k_1 = 1 }^{L}  q^{ ( a_1 + (a_3-a_2) - a_3 )  
(  k_1 - \cF ( k_1, 0  ~;~ \vec i )  ) }  \cr 
& = \sum_{k_1 = 1 }^{L}  q^{ ( a_1 - a_2 )  
(  k_1 - \cF ( k_1, 0  ~;~  \vec i )  ) }  \cr 
&= \cS ( a_1 - a_2 ~;~ \vec i  ) \cr }}
This sum is again of the 
same form  as \ktsm\ except that $a_2$ is now replaced 
by $a_1 -a_2$ and $ \cF (k_2, k_1 ~;~ \vec i  )$  by 
 $ \cF (k_1, 0 ~;~ \vec i  )$. 
Combining the result of the sums 
we get 
\eqn\finthrt{\eqalign{  
& TR ~(~ \tau_{a_3, 0 }  \tau_{a_2, 0 } \tau_{a_1,0 } ~ 
\Delta_q (E_{21}^n ~ )~~   \Phi_1^{ \otimes L }   ~)~ \cr 
& =
q^ { (L+1) n \ovt }[n]_q!~~ \sum_{1 \le i_1  < i_2 \cdots < i_n  \le   L } 
q^ { - ( i_1 + i_2 + \cdots i_n)  }~~~ TR  ~~\bigl (~~ \rho_{i_1 } ( E_{21} )
\cdots  \rho_{i_n } ( E_{21} )  ~~~ \Phi_1^{\otimes L } ~~\bigr ) ~~  \cr
&\qquad \qquad \qquad \qquad \qquad\qquad \qquad \cS ( a_1-a_2 ~;~ \vec i  ) 
\cS  ( a_2-a_3 ~;~ \vec i ) \cS ( a_3 ~;~ \vec i  ) \cr }}
where $\cS$ was defined in \defS. 

In the case of $P$ $\tau$'s the same kinds of manipulations lead to 
\eqn\finthrt{\eqalign{  
& TR ~(~ \tau_{a_P, 0 }  \cdots \tau_{a_2, 0 } \tau_{a_1,0 } ~ 
\Delta_q (E_{21}^n ~ )~~   \Phi_1^{ \otimes L }   ~)~ \cr 
& =
q^ { (L+1) n \ovt }[n]_q!~~ \sum_{1 \le i_1  < i_2 \cdots < i_n  \le   L } 
q^ { - ( i_1 + i_2 + \cdots i_n)  }~~~ TR  ~~\bigl (~~ \rho_{i_1 } ( E_{21} )
\cdots  \rho_{i_n } ( E_{21} )  ~~~ \Phi_1^{\otimes L } ~~\bigr ) ~~  \cr
&\qquad \qquad \qquad \qquad \qquad
 \cS ( a_1-a_2 ~;~ \vec i  ) 
\cdots   \cS  ( a_{p-1} -a_p ~;~ \vec i ) \cS ( a_p ~;~ \vec i  ).}}

The final step is to rewrite this operator in a form that can
be compared to the BMN operators given in \BMNtwo.  The idea
is exactly as described in the two $\tau$ case in the discussion of
formulae \resttdtwo\ to \fintwotau.  Using the definition
\defnT\ we find
\eqn\finPtau{ \eqalign{ 
& TR ~(~ \tau_{a_P, 0 }  \cdots  \tau_{a_2, 0 } \tau_{a_1,0 } ~ 
\Delta_q (E_{21}^n ~ )~~
  \Phi_1^{ \otimes L }   ) \cr   
& =  { (-1)^n  \over 1 - q^{a_1 - n}} [n]_q!~~ 
\sum_{0 \le k_1  \le k_2 \cdots \le
k_{n-1}  \le   J } \left( 1 - {1 \over n-1} \sum_{l=1}^{n-1} 
q^{(a_1 - n)(1-k_l)}
\right) \cr
& q^ { - ( k_1 + k_2 + \cdots k_{n-1} )  }~
\cT ( a_1 -a_2 ~;~ \vec k  ) \cdots \cT ( a_{P-1} -a_P ~;~ \vec k  )  
\cT ( a_P ~;~ \vec k ) \cr
& TR [, k_1, k_2 - k_1,...,k_{n-1} - k_{n-2}, J - k_{n-1}]. }}
This is our final expression for the trace which uses 
$P$ $\tau$ operators.
Comparing to \BMNtwo\ one sees that it is a linear combination
of many BMN operators.  However, it is important to note
that in this combination there is only one occurence of an 
operator with the most generic oscillator numbers,
in this case $P$ generic oscillator numbers.  It is of the
form
\eqn\Pcompare{\eqalign{ 
TR ~(~ \tau_{a_P, 0 }  \cdots  \tau_{a_2, 0 } \tau_{a_1,0 } ~ 
& \Delta_q (E_{21}^n ~ )~
\Phi_1^{ \otimes L }   ) \leftrightarrow \cr   
& \alpha^{\dagger}_{n-1-a_1} (\alpha^{\dagger}_{-1})^{n-P-1} 
\alpha^{\dagger}_{a_P -1} \prod_{l=1}^{P-1} \alpha^{\dagger}_{a_l - 
a_{l+1} - 1}
+ \cdots}}
where the $\cdots$ denotes operators with $n-P$ or more 
$\alpha^{\dagger}_{-1}$'s.
It is straightforward in practice therefore to take linear combinations of
the $P$ $\tau$ operator with lower $\tau$ operators to isolate this
generic oscillator number string state.  Moreover by taking $P=n-1$ one
can obtain the most general oscillator number state involving insertions
of $\Phi_2$ operators.


\newsec{ Implications of the quantum group construction of BMN operators  } 

We discuss some physical and mathematical implications of the 
quantum group construction of  BMN operators presented 
in the earlier sections.

\subsec{  Correlators as traces of quantum group generators } 

Since we have expressed the BMN operators 
in terms of the co-product of the quantum group 
and $q$-cyclic operators built from quantum group 
generators, we may expect that we should be able 
to express the correlators of BMN operators in terms 
of traces of quantum group generators and $\tau $ operators 
acting on tensor space. For operators in half BPS 
representations, this step of expressing correlators 
as traces of group theoretic quantities in tensor space 
was described in \refs{\excor,\finfac } and was used 
to derive factorization equations and to exhibit 
relations betwen correlators in the 
four dimensional theory and  classical ( large $k$ ) Chern Simons
theory. We outline some steps in this direction for BMN operators. 

Let us focus on the case which has been the main focus of the 
previous section, namely where the impurities 
are all one complex $ \Phi_2$. Both $ \Phi_1 $ and $ \Phi_2 $ 
are matrices which transform an $N$-dimensional 
space $ V$. It is useful to consider an operator which 
collects both of them into one object. In a sense we are thinking 
of a $U(N)$ theory with two flavors as a $U(2N)$ theory 
broken to $ U(N)$.  We  define 
a matrix  $ \Phi = \Phi_1 \oplus \Phi_2 $ or in matrix notation 
\eqn\bigmat{ 
\Phi =   \pmatrix {   \Phi_1 &  0 \cr 
                      0     &  \Phi_2 \cr }
}
which acts on 
two copies of $V$, i.e $W= V \oplus V $. 
Projection  projectors $ P_1 , P_2$ 
\eqn\proj{\eqalign{  
& P_1 = \pmatrix {   1 &  0 \cr 
                   0 &  0   \cr }  \cr 
& P_2 = \pmatrix { 0  &  0 \cr 
                   0 &   1  \cr } \cr }}
project to the first and second copy 
of $V$ respectively. This allows us to 
write $ \Phi_1 = \Phi P_1 $ and $ \Phi_2 = \Phi P_2 $. 
The basic  free field 
two point functions of $ \Phi_1 $ with $ \Phi_1^{\dagger}$, 
and of $ \Phi_2 $ with $ \Phi_2^{\dagger} $ and 
the vanishing of the two-point function of 
$ \Phi_1$ with $ \Phi_2^{\dagger} $ or $ \Phi_2$ with $ \Phi_1^{\dagger} $
are all encoded in the formula 
\eqn\bastwopt{ 
 \langle \Phi \Phi^{\dagger} \rangle  =  
( P_1 \otimes P_1 + P_2 \otimes P_2 )\circ \gamma   } 
where the $ \Phi $ and $ \Phi^{\dagger } $ are viewed as 
operators in $ W \otimes W $ and $\gamma $ is a twist 
 which permutes one $ W$ with the other. 
More generally we think of $ \Phi$ and $ \Phi^{\dagger}$ each 
as  operators acting on the tensor product $ W^{\otimes n }$, 
and the two-point function can be written as a sum 
of insertions of $P_1 \otimes P_1 $ and $ P_2 \otimes P_2$. 
The evaluation of correlation functions can then be mapped, 
using formulas we have developed, into the evaluation of traces 
of sequences of operators. The operators will include 
the $P_1,P_2$ projectors, as well as the quantum 
group generators $ \Delta_q( E_{21} ) $, $\Delta_q ( E_{12} ) $ 
and the $q$-cyclic operators.

In \refs{ \excor,\finfac }  the operators involved
after evaluating the two-point functions of the Higgs fields
were all permutations, essentially because a general multitrace 
highest weight half-BPS operator could be written as $ tr ( \sigma
\Phi_1 ) $, or as $ tr ( P_R  \Phi_1 ) $ where  $ \Phi_1$ 
 was defined to act in $ V^{\otimes n } $ and the trace 
was taken in the $n$-fold tensor product. $P_R$ is a projection 
operator onto  Young Diagrams, and orthogonality 
of these projectors allowed one to diagonalize the two-point
functions. 
The diagonalization of  BMN operators is now a question related 
to finding projection operators in tensor space $ W^{\otimes n }$.  
This new perspective on the BMN operators should be useful 
in further studies of their correlators. It is interesting that 
the expression of half BPS operators in the form 
$ tr ( \sigma \Phi_1 ) $ also plays a role in 
the string bit model \verl\vaverl. 

One of the interesting features of physical applications of 
quantum groups at roots of unity is that they capture vanishing properties 
of correlation functions or fusion rules \pasqsal\mose\alvgom.  
It will be interesting to look for signatures 
of such vanishings in this context. 
For example, in all the calculations of section 
$5$, BMN operators emerge from the quantum group 
construction with the $q$-factorial $[n]_q!$ which vanishes 
at $ q^n = e^{ 2 i n  \pi \over J } =  \pm  1 $. So at $ n = J $ 
( and $ n = J/2$ for $J$ even ),  
we have a qualitative change from the point of view of the 
quantum group construction. It will be interesting to see if this 
is reflected in the correlators computed either from the super-Yang
Mills or the string field theory.

\subsec{Remark on the quantum group transformation of the 
action } 

It is interesting to ask if the type of quantum 
deformation of the global symmetry group of 
SYM can be given meaning as a transformation 
of the action. We do not expect it to be a symmetry
since it is a spectrum generating algebra
which does not commute with the Hamiltonian. 
But we would like to see if a consistent 
definition can be given of the transformation rule. 
We will not  explore this in detail here, except to 
indicate that a well-defined transformation is indeed 
possible. Consider for example the term 
\eqn\action{
\int d^4 x TR  ([ \Phi_1 , \Pd_1 ] + [ \Phi_2, \Pd_2])^2 
} 
in the action. 
Expanding it out one finds many terms.
There is a well defined action of the quantum group 
using the quantum co-product on a product 
of $\Phi$'s ( the algebra of $ \Phi$'s can be given 
the structure of a module algebra, as defined for example 
in \charpress\ ). But we have traces, which only 
determine a product up to cyclicity. 
We can use the cyclicity to write the trace in a manifestly
cyclic symmetric form and then act on the sequence 
of products thus obtained. 
For example, consider the term in the expansion of the
operator \action\ above, 
\eqn\cycsym{ 
TR ( \Phi_1 \Pd_1 \Phi_2 \Pd_2 ) 
= {1 \over 4 }TR ( \Phi_1 \Pd_1 \Phi_2 \Pd_2  +
\Pd_1 \Phi_2 \Pd_2 \Phi_1 +
\Phi_2 \Pd_2 \Phi_1 \Pd_1 +
\Pd_2 \Phi_1 \Pd_1 \Phi_2 ).}
Applying this procedure to all the terms appearing in the
expansion of \action\ produces many more terms. 
Now we can act on each term appearing in this expansion 
using the quantum co-product. 
For example the action of $E_{21} $ on  $ \Phi_1 \Pd_1 \Phi_2 \Pd_2$ 
gives 
\eqn\exampleaction{
{1 \over 2} (q^{-3/2} + 2 q^{-1/2} + q^{1/2}) TR( \Pd_1 \Phi_2
\Pd_2 \Phi_2) -
{1 \over 2} (q^{-1/2} + 2 q^{1/2} + q^{3/2}) TR( \Phi_2 \Pd_1
\Phi_1 \Pd_1)}
where we have applied the trace in arriving at this form.
Computing the action of $E_{21}$ on all the terms in the expansion
of \action,
we find
\eqn\anstrans{\eqalign{ 
{1 \over 2} \big(&  (- q^{-3/2} + q^{-1/2} + q^{1/2} - q^{3/2})
TR ( \Phi_2 \Pd_1 \Phi_1 \Pd_1) \cr & +
(q^{-1/2} - q^{3/2}) TR ( \Phi_2 \Phi_1 (\Pd_1)^2)
+ (- q^{-3/2} + q^{1/2}) TR ( \Phi_2 (\Pd_1)^2 \Phi_1) \cr
& + ( q^{-3/2} - q^{-1/2} - q^{1/2} + q^{3/2})
TR ( \Pd_1 \Phi_2 \Pd_2 \Phi_2) \cr & +
(- q^{-1/2} + q^{3/2}) TR ( \Pd_1 \Pd_2 (\Phi_2)^2) +
(q^{-3/2} - q^{1/2}) TR ( \Pd_1 (\Phi_2)^2 \Pd_2) \big).
}} 
So the action is not invariant (although it becomes invariant
as $q \rightarrow 1$ as is easily checked in the term above) 
but transforms in a
specified way. 
It will be interesting to see if Ward identities 
can be developed using these transformations of the action, 
and if they have useful information for correlators 
of BMN operators.

\subsec{  Quantum  group symmetry and quantum geometries } 

It is tempting to conjecture that the quantum group 
construction has a geometrical meaning in terms of quantum spaces.
While we have explicitly shown the construction of 
BMN operators with correct symmetry using the class 
of single impurity insertions generated by an $U_q ( U(2)  ) $ subgroup, 
many of our considerations should apply to 
the construction of the most general operators 
using the full $q$-deformed superalgeba $SU_q(4|2,2 )$. 
We have also commented that the 
$q$-deformed superalgebra $ SU_q ( 3 |2,1 )$ 
can be expected to play a special role related to holomorphic 
insertions. Suggestions that $SU_q(4|2,2 )$
 might be relevant to $N= 4$ SYM were made in the 
 context of a conjecture that quantum $ ADS \times S $  spacetimes 
are relevant to finite $N$ effects \jr\hrt . The $ q $ in those discussions 
was also a root of unity, but a different one  $ q=e^{2 \pi i \over
N}$, chosen to capture certain truncations in the spectrum of chiral
primaries associated with finite N. These truncations are related to
the stringy exclusion principle \malstrom\ and giant gravitons\mst.
In this context, $q$-deformed spectrum generating algebras
have also been discussed \bgm. 
The  exploration of the connection between the 
algebraic constructions here and the stringy exclusion principle, 
giant gravitons and non-commutative 
spacetimes is an interesting problem we leave for the future. 
  
The idea that there is some geometrical meaning 
to the quantum group construction of the BMN operators 
is  also suggested by the technical similarities between 
the $q$-cyclic operators we have used and analogous operators 
that appear in cyclic cohomology of quantum groups. 
For example  in \conmos\ a map is found  between  cyclic cohomology 
of quantum groups and that for   module algebras 
which are equipped with $q$-cyclic traces. 
Finding  concrete connections between the work of \conmos\ 
and the work of BMN is a fascinating direction. 
 At least some known properties of cyclic cohomology 
 may be taken to suggest the existence of
 a connection to  the physical context of
 strings on plane waves. For example, Connes shows 
 that the cyclic cohomology of the group ring 
 of a finite group $ \Gamma $  is related to the $S^1$ equivariant 
cohomology  of the loop space related to the classifying 
space of $ \Gamma$ ( see section 2.$\gamma$ of  \connes). 
Assuming analogous results exist 
for the cyclic cohomology of quantum groups at roots of unity,
 the appearance of 
loop spaces would be mirrored on the physical side 
by presence of maps from a string. The $S^1$ 
equivariance is suggestive of residual diffeomorphism invariance. 
The appearance of the classifying space of the quantum group 
is suggestive of quantum homogeneous spaces  
which include   quantum deformations of  
 $ADS \times S$ or of the pp-wave background.
Finding a more concrete formulation of 
ideas in this direction would be interesting, especially 
since  they may  give  insight into the quantum 
geometrical  meaning of correlators  of  stringy states  in a
plane wave background.

\bigskip

\noindent{ \bf Acknowledgements: }
 We wish to thank for discussions 
 Antal Jevicki, David Lowe, Horatiu Nastase. 
 This research was supported  by DOE grant  DE-FG02/19ER40688-(Task A).

\bigskip
\bigskip

\noindent

\noindent

{\bf Appendix  A :   Constructing BMN operators}

Motivated by the discussions in \refs{\constab,\mjr}, 
we construct the BMN
operators in the following way.  First we define
an intermediate set of fields
\eqn\tildefields{{\tilde \Phi}_{\beta,p,k} \equiv 
\Phi_{1}^{k} \Phi_{\beta} \Phi_{1}^{-k} q^{k p}}
The BMN operators can then be constructed as
\eqn\BMNopstwo{{\cal O}_{\beta_n,p_n; \beta_1, p_1, \dots ,
\beta_{n-1}, p_{n-1} } =
{\cal N}_n \sum_{0 \leq i_1 \leq i_2 \leq \cdots \leq i_n \leq J} ~
\sum_{\sigma \in S_n} TR  \left(
{\tilde \Phi}_{\beta_{\sigma (1)},p_{\sigma (1)},i_1} \cdots 
{\tilde \Phi}_{\beta_{\sigma (n)},p_{\sigma (n)},i_n}
\Phi_{1}^{J} \right)}
This expression  treats
all ${\tilde \Phi}$ fields symmetrically, and
as such it provides a simple generalization of the original
BMN prescription for the two-impurity case 
\refs{\bmn,\constab,\kpss}.
We must however show that this operator
reduces to that of the two-impurity case when $n=2$.
Indeed, the two-impurity case as originally
constructed in \bmn\ comes with
only one summation index, whereas the form
proposed above comes with two.  Our task
in this appendix is to show that \BMNopstwo\
does indeed reduce to the more familiar form
given in section 2.

We begin by rewriting the operator \BMNopstwo\ in terms of
the notation \opnotation\ introduced in section 2.  We find
\eqn\BMNopsthree{\eqalign{ {\cal O}_{\beta_n,p_n ; \beta_1, p_1 , \cdots ,
\beta_{n-1}, p_{n-1}  } = &
{\cal N}_n \sum_{0 \leq i_1 \leq i_2 \leq \cdots \leq i_n \leq J} ~
\sum_{\sigma \in S_n}  q^{\sum_{l=1}^{n} i_l p_{\sigma(l)}}
 TR  [i_1 {\beta_{\sigma (1) } \atop ,} i_2 - i_1 \cr &
{\beta_{\sigma (2) } \atop ,} i_3 - i_2  
{\beta_{\sigma (3) } \atop ,} \cdots 
{\beta_{\sigma (n) } \atop ,} J - i_n ] .}}
This form can be simplified somewhat by using cyclicity
of the trace and redefining summation variables.
In particular one notes that cycling the term
$i_1$ in the trace to the end of that operator
produces the term $J + i_1 - i_n$.  Therefore
there are only $n-1$ different variables in the
operator that are being summed over.  Consequently
one of the sums can be done explicitly.  One makes
this manifest in the following way.  For any given 
permutation $\sigma \in S_n$, we use cyclicity of
the trace to cycle the $\beta_n$ operator insertion
to the first position.  For example, if $\sigma (s) = n$,
then we cycle the operator into the form
\eqn\cycle{
TR  [ {\beta_n \atop ,} i_{s+1} - i_s {\beta_{\sigma (s+1)} \atop ,}
i_{s+2} - i_{s+1}  {\beta_{\sigma (s+2)} \atop ,} \cdots
{\beta_{\sigma (n) } \atop ,} J + i_1 - i_n {\beta_{\sigma (1) } \atop ,}
\cdots {\beta_{\sigma (s-1) } \atop ,} i_{s} - i_{s-1} -1].}
This form of the operator suggests redefining the summation
indices in the following way:
\eqn\redefn{\eqalign{k_1 & = i_{s+1} - i_s  \cr
k_{l}-k_{l-1} & = i_{s+l} - i_{s+l-1}, ~~ 2 \leq l \leq (n-s) \cr
k_{n-s+1} - k_{n-s} & = J + i_1 - i_n  \cr
k_{l} - k_{l-1} & = i_{l-(n-s)} - i_{l - (n-s) -1}, ~~
n-s+2 \leq l \leq n-1 .}}
Equivalently one can solve for the $k_l$'s as
\eqn\solvek{\eqalign{k_l & = i_{s+l} - i_s , ~~ 1 \leq l \leq n-s
\cr
k_l & = J + i_{l-(n-s)} - i_s , ~~ n-s+1 \leq l \leq n-1.}}
The new set of summation variables now consists of
$i_s$ and the $k_l$'s for $1 \leq l \leq n-1$.
In terms of these variables the  operator \cycle\ becomes
\eqn\cycletwo{
TR  [ {\beta_n \atop ,} k_1 {\beta_{\sigma (s+1)} \atop ,}
k_{2} - k_{1}  {\beta_{\sigma (s+2)} \atop ,} \cdots
{\beta_{\sigma (n) } \atop ,} k_{n-s+1} - k_{n-s} {\beta_{\sigma (1) } \atop ,}
\cdots {\beta_{\sigma (s-1) } \atop ,} J - k_{n-1}].}
In particular the $i_s$ dependence drops out of the operator.

To do the $i_s$ sum we make note of the following facts.
First the sum over the permutation group $S_n$ can
be rewritten as 
\eqn\permsum{\sum_{\sigma \in S_n} = \sum_{s=1}^{n} ~~
\sum_{\tau \in S_{n-1}}}
That is, given a permutation $\sigma \in S_n$ satisfying
$\sigma (s) = n$, we can construct a permutation
$\tau \in S_{n-1}$ satisfying 
\eqn\tauconstruct{\eqalign{\tau (l) & = \sigma (s+l), ~~ 1 \leq l \leq n-s
\cr
& = \sigma (l-n+s), ~~ n-s+1 \leq l \leq n-1}}
For a given $s$, the $S_{n-1}$ sum is simply over
all $\tau$ constructed in this way.  The sum
on $s$ then fills out the remaining elements of
the $S_n$ permutation group.
Secondly we note that the sums over the $i_l$ indices becomes
\eqn\indexsum{\sum_{0 \leq i_1 \leq i_2 \leq \cdots \leq i_n \leq J}
= \sum_{0 \leq k_1 \leq k_2 \leq \cdots \leq k_{n-1} \leq J}
~~ \sum_{i_s = J-k_{n-s+1}}^{J-k_{n-s}}}
in the $k_l$ indices.
This follows simply from the index redefinitions given
in \solvek.  
The last fact that we need is to rewrite
the phase factor as
\eqn\phasere{\eqalign{q^{\sum_{l=1}^{n} i_l p_{\sigma(l)}}
= q^{\sum_{l=1}^{n} i_{\sigma^{-1} (l)} p_{l}} 
= q^{\sum_{l=1}^{n-1} k_{\tau^{-1}(l)} p_{l} + i_s (p_1 + \cdots p_n)}
& =q^{\sum_{l=1}^{n-1} k_l p_{\tau(l)} + i_s (p_1 + \cdots p_n)}.}}
The first and third equality signs follow trivially.  The second equality
follows from the definition of $\tau$ given in \tauconstruct\
and the assumption that $q^J = 1$.
If we take $p_1 + \cdots p_n =0$, as one usually does
to obtain an operator that corresponds to a string
state, then all dependence on the summation index
$i_s$ drops out.  However we shall not make this
assumption here.  As we shall see in a moment, keeping
all $p_l$'s generic merely results on the string
side to considering a string state which is a linear
superposition of different ($n$ for generic $p_l$'s)
single string states.

Now we combine these basic facts to reproduce the
operator given in section 2.  From the permutation
sum, the $i_l$ sums, and the $q$-factor discussed
in the preceding paragraph, all $s$
and $i_s$ dependent factors reduce to
\eqn\extrastuff{\eqalign{\sum_{s=1}^{n} ~ \sum_{i_s = J-k_{n-s+1}}^{J-k_{n-s}}
q^{i_s (p_1 + \cdots + p_n)} 
}}
where we have introduced the new $k$ indices
$k_0$ and $k_n$ which are fixed to $0$ and $J$ respectively.
These arise from the special cases $\sigma (n) = n$ and
$\sigma (1) = n$ respectively and are easily checked
to have the values just quoted.
This sum is straightforward to evaluate.  
The $s$ and $i_s$ sums together fill
out a sum (replacing $i_s$ by $i$) $\sum_{i=0}^{J}$,
but neighboring $i_s$ sums overlap exactly at the
endpoints, therefore \extrastuff\ can be rewritten as
\eqn\extratwo{\sum_{i=0}^{J} q^{j (p_1 + \cdots + p_n)}
+ \sum_{l=1}^{n-1} q^{(J-k_l) (p_1 + \cdots + p_n)}.}
Depending on whether $p_1 + \cdots + p_n$ vanishes or not,
this sum evaluates to
\eqn\extrathree{\eqalign{&= 1 +  \sum_{l=1}^{n-1} 
q^{-k_l  (p_1 + \cdots + p_n)}, ~ p_1 + \cdots + p_n \neq 0 \cr
& = J + n, ~ p_1 + \cdots + p_n = 0.}}

Putting everything together reproduces the
operator \BMNfinal\ given in section 2 provided
that $p_1 + \cdots + p_n = 0$ and where
the factor of $J+n$ in \extrathree\ has been
absorbed into the normalization.  If instead
we take  $p_1 + \cdots + p_n \neq 0$ we obtain the
operator
\eqn\BMNnonzero{\eqalign{
& {\cal O}_{\beta_n,p_n ; \beta_1, p_1 , \cdots \beta_{n-1}
, p_{n-1} } =  {\cal N}_n
\sum_{0 \leq k_1 \leq k_2 \leq \cdots \leq k_{n-1} \leq J}
~ \sum_{\tau \in S_{n-1}}  \cr 
& q^{\sum_{l=1}^{n-1} k_l p_{\tau (l)}}
(1 +  \sum_{l=1}^{n-1} 
q^{-k_l (p_1 + \cdots + p_n)}) 
 TR  [ {\beta_n \atop ,} k_1 {\beta_{\tau (1)} \atop ,} k_2 - k_1
{\beta_{\tau (2)} \atop ,} \cdots {\beta_{\tau (n-1)} \atop ,}
J - k_{n-1} ] }}
where now an extra sum of $q$-factors appears
as compared to \BMNfinal.  The intepretation
however is clear.  This operator corresponds not
to a single string state, but rather to a linear
superposition of string states.  Explicitly we find
the correspondence 
\eqn\BMNdualtwo{\eqalign{{\cal O}_{\beta_n,p_n ; \beta_1, p_1, \dots
,\beta_{n-1} p_{n-1} }
\leftrightarrow
\sum_{l=1}^{n} \alpha^{\dagger}_{-(p_1 + \cdots + p_n) + p_l} 
\prod_{k=1, k \neq l}^{n-1} \alpha^{\dagger}_{p_k}}}
Such a formula is expected as the starting point \BMNopsthree\
treats the $n$ oscillator numbers $p_l$ symmetrically.

Finally we would like to emphasize the importance of the
sum over the permutation group $S_n$ in
\BMNopstwo\ in
comparing to the dual string state in \BMNdual\ or
\BMNdualtwo.
The creation operators in \BMNdual\ and \BMNdualtwo\ commute;
therefore, this property should be evident in
the dual operator.  
The sum over the permutation group  makes
this property manifest, and moreover is necessary
for the correspondence in \BMNdual\ and \BMNdualtwo\ to make sense.
In the two-impurity case, this property is satisfied
trivially, and as such was not an issue in 
\bmn.   Already at the three-impurity level
however this sum is necessary as was noted in
\constab\ as part of a construction for $n$-impurities. 
The $n$-impurity case was discussed further in \mjr.

\bigskip 

{ \bf Appendix  B  : The degenerate case $a_1 =n $  }
\medskip 

In section 5.2 we described how to get BMN 
operators using  
$ TR ( \tau_{a_1} ( \Delta ( E_{21}^n ) \Phi_1^{J+n} ) $. 
In doing the sums we assumed that $a_1 \ne n $. 
The case $a_1 = n $ has to be treated separately. 
Specializing \simpttdp\ to the case $a_1 = n$ gives the result
\eqn\pfsto{\eqalign{ 
&  TR \tau_n \Delta_q ( E_{21}^n ) [ J+n ] 
 =   (-1)^{n } [n]_q !   \sum_{0 \le j_1 \le j_2 \cdots j_n \le J } 
 (  q^{n j_1} + q^{n j_2 } + \cdots + q^{n j_n} ) \cr 
&  \qquad \qquad TR  [j_1 , j_2-j_1 , \cdots , j_n - j_{n-1}  , J - j_n ]  
   ~~~ q^{ -j_1 -j_2 \cdots j_n  } \cr }}

We cycle the first $j_1$ operators to the right
and  change variables 
\eqn\chngjk{\eqalign{  
& k_1 = j_2-j_1 \cr  
& k_2 - k_1 = j_3-j_2  \cr 
& \qquad \vdots \cr 
& k_{n-1} = j_n - j_{n-1}  \cr }}
which implies 
\eqn\chngjki{\eqalign{  
& k_1 = j_2-j_1 \cr  
& k_2  = j_3-j_1  \cr 
& \qquad  \vdots \cr 
& k_{n-1} = j_n - j_1 \cr }}
We now write the sum in terms of the variables $ ( j_1 ; k_1 \cdots
k_{n-1} ) $.  
We write  the sum in \pfsto\ as a sum of $n$ copies of the same thing
with an overall factor of $1/n$. 
 The first term is
\eqn\spltsum{\eqalign{  
& (-1)^{n} [n]_q !  \sum_{0  \le k_1 \cdots  k_{n-1}
 \le J }  \sum_{ j_1 =0}^{J- k_{n-1}  }
    TR  ( [ , k_1  , k_2-k_1,  \cdots , k_{n-1}  -k_{n-2} , 
J - k_{n-1 } ]  ) \cr 
& \qquad\qquad\qquad\qquad\qquad
 q^{ - k_1  \cdots - k_{n-1} }  
( 1 + q^{ n k_1} + q^{nk_2} + \cdots q^{nk_{n-1} } )
\cr 
& =  (-1)^{n} [n]_q !   
\sum_{0  \le k_1 \le k_2 \cdots \le  k_{n-1} \le J } 
TR ( [ , k_1  , k_2 -k_1, \cdots , k_{n-1 -k_{n-2}} ,  J - k_{n-1}  ])
\cr
& ( J - k_{n-1} + 1 ) ( 1 + q^{ n k_1} + q^{nk_2} + \cdots q^{nk_{n-1} } )
  q^{ (   -k_1 - k_2 - \cdots k_{n-1}  )  } \cr 
 }}

In the second term we will  cycle 
one $\phi_2$ impurity as well to get 
\eqn\sttwo{\eqalign{  
& (-1)^{n} [n]_q ! 
\sum_{0  \le k_1 \le k_2 \cdots \le  k_{n-1} \le J } 
 \sum_{ j_1=0}^{J-k_{n-1}  }
 TR  ( [ , k_2 - k_1 , k_3- k_2 , \cdots , k_{n-1} - k_{n-2} ,
J-k_{n-1} , k_1    ] \cr 
& \qquad\qquad\qquad ( 1 + q^{ n k_1} + q^{nk_2} + \cdots q^{nk_{n-1} } )
q^{ (   -k_1 - k_2 - \cdots k_{n-1}  )  }
\cr }}
Now we do a relabelling 
\eqn\fstrlab{\eqalign{  
&  \tilde{k}_{1}  = k_2 -k_1  \cr 
&  \tilde{k_2} - \tilde{k_1} = k_3 - k_2  \cr 
& \qquad \vdots \cr 
& \tilde{k}_{n-2} - \tilde k_{n-3}    = k_{n-1} - k_{n-2}  \cr  
& J -  \tilde{k}_{n-1}  = k_1 \cr }}
to get  operator in the summand to be of the same form 
as \spltsum. 

This leads the $q$ factors to transform as 
\eqn\fstrq{\eqalign{  
& q^{ -k_1 - \cdots - k_{n-1} } = q^{ n\tilde k_{n-1} } q^{ - \tilde k_1
- \cdots  - \tilde k_{n-1} } \cr 
& ( 1 + q^{ n k_1} + \cdots + q^{ n k_{n-1} } ) 
  = q^{ - n \tilde k_{n-1} } ( 1 + q^{ n \tilde k_1} + \cdots + q^{ n
\tilde k_{n-1} } )  \cr }}
which implies 
\eqn\fstrqi{ q^{ -k_1 - \cdots - k_{n-1} }( 1 + q^{ n k_1} + \cdots +
 q^{ n
k_{n-1} } ) 
  = q^{ - \tilde k_1
- \cdots  - \tilde k_{n-1} }
( 1 + q^{ n \tilde k_1} + \cdots + q^{ n \tilde k_{n-1} } )     }  
Thus the $q$ factors as well as the 
operator are  identical in the tilde variables
as can be seen by comparing to  \spltsum. The only difference is that
the coefficient is $ ( \tilde k_{n-1}- \tilde k_{n-2  } +1 ) $. 
After renaming the tilded variables back to untilded 
variables and collecting the first two terms 
we get a coefficient $ ( J -k_{n-1} + 1 ) + (k_{n-1} - k_{n-2} + 1 ) $.  
Another cycling step produces a sum of the same form 
with the coefficient of $ ( k_{n-2} - k_{n-3} +1 ) $. 
Continuing this procedure and collecting terms we get
$ ( J - k_1 + 1 ) + (k_{n-1} - k_{n-2} + 1 ) +  ( k_{n-2} - k_{n-3}
+1 ) + ( k_{n-3} - k_{n-4} + 1 ) + \cdots + (k_{2} - k_1 + 1 ) + ( k_1
+1 ) = ( J +n ) $. This leads to the result 
 \eqn\res{\eqalign{ 
& TR   \tau_{n}  
 (  \Delta_q ( E_{21}^n   ) \Phi_1^{\otimes (J + n) } ) \cr
& = { J+n \over n } [n]_q! 
 \sum_{ 0 \le k_1 \le k_2 \cdots k_{n-1}  \le  J } 
TR  ( \Phi_2 \Phi_1^{ k_1 } \Phi_2 
\Phi_1^{ k_2 -k_1 } \Phi_2 \cdots \Phi_2 \Phi_1^{ k_{n-1} - k_{n-2} } 
 \Phi_2 \Phi_1^{ J - k_{n-1} }  ) ~~ \cr 
&  q^{- k_1 - \cdots - k_{n-1} }   ( 1 + q^{ n k_1} + \cdots + 
q^{ n k_{n-1} } ) \cr 
& \rightarrow \alpha^{(2)}_{(n-1) } \alpha^{(2)}_{-1} \cdots
\alpha^{(2)}_{-1} \alpha^{(2)}_{-1} |0 > \cr
}}

\listrefs

\end